\documentclass[prd,twocolumn,twoside,preprintnumbers,superscriptaddress,nofootinbib,floatfix]{revtex4}

\usepackage{amsmath}
\usepackage{amsfonts}
\usepackage{graphicx}
\usepackage{multirow}
\usepackage{xcolor}
\usepackage{hyperref}
\usepackage{float}
\usepackage{ulem}

\def\be{\begin{equation}}
\def\ee{\end{equation}}
\def\beq{\begin{equation}\begin{aligned}}
\def\eeq{\end{aligned}\end{equation}}

\newcommand{\Eq}[1]{Eq.~\eqref{#1}}

\def\tev{\, {\rm TeV}}
\def\gev{\, {\rm GeV}}
\def\mev{\, {\rm MeV}}

\newcommand{\ZZ}{\mathbb{Z}}
\newcommand{\absval}[1]{\left| #1 \right|}

\newcommand{\bs}[1]{\boldsymbol #1}

\begin{document}

\preprint{\vbox{\hbox{SCIPP 17/12}}}

\title{Dark Matter Freeze-in Production in Fast-Expanding Universes}

\author{Francesco~D'Eramo}
\email{francesco.deramo@pd.infn.it}
\affiliation{Dipartimento di Fisica ed Astronomia, Universit\`a di Padova, Via Marzolo 8, 35131 Padova, Italy}
\affiliation{INFN, Sezione di Padova, Via Marzolo 8, 35131 Padova, Italy}

\author{Nicolas~Fernandez}
\email{nfernan2@ucsc.edu}
\affiliation{Department of Physics, 1156 High St., University of California Santa Cruz, Santa Cruz, CA 95064, USA}
\affiliation{Santa Cruz Institute for Particle Physics, 1156 High St., Santa Cruz, CA 95064, USA}

\author{Stefano~Profumo}
\email{profumo@ucsc.edu}
\affiliation{Department of Physics, 1156 High St., University of California Santa Cruz, Santa Cruz, CA 95064, USA}
\affiliation{Santa Cruz Institute for Particle Physics, 1156 High St., Santa Cruz, CA 95064, USA}

\date{December 20, 2017}
\begin{abstract}
If the dark matter is produced in the early universe prior to Big Bang nucleosynthesis, a modified cosmological history can drastically affect the abundance of relic dark matter particles. Here, we assume that an additional species to radiation dominates at early times, causing the expansion rate at a given temperature to be larger than in the standard radiation-dominated case. We demonstrate that, if this is the case, dark matter production via freeze-in (a scenario when dark matter interacts very weakly, and is dumped in the early universe out of equilibrium by decay or scattering processes involving particles in the thermal bath) is dramatically suppressed. We illustrate and quantitatively and analytically study this phenomenon for three different paradigmatic classes of freeze-in scenarios. For the frozen-in dark matter abundance to be as large as observations, couplings between the dark matter and visible-sector particles must be enhanced by several orders of magnitude. This sheds some optimistic prospects for the otherwise dire experimental and observational outlook of detecting dark matter produced by freeze-in.

 \end{abstract}
%

\maketitle

\section{Introduction}

Thermal freeze-out is an attractive mechanism for dark matter (DM) genesis~\cite{Lee:1977ua,Scherrer:1985zt,Srednicki:1988ce,Gondolo:1990dk}. Within this paradigm, DM particles are in thermal equilibrium at high temperatures; as the plasma temperature eventually drops below the DM mass, the Hubble expansion rate becomes larger than the rate for processes that keep the DM species in thermal equilibrium; DM particles thus {\it freeze-out},  with an approximately fixed comoving number density. Remarkably, such a relic density depends only on masses and couplings that can be, in principle, independently measured in a laboratory, and it therefore does not depend on the uncertain cosmological history of the universe. The jargon used to express this fact is to say that DM freeze-out is ``{\it IR-dominated}''. 

The above statement has, however, a well-known caveat: it is true only for a standard thermal history (i.e. an energy density dominated by radiation at early times, $T\gg 1 \, {\rm MeV}$) all the way up to the freeze-out temperature, approximately a factor of 20 below the DM mass. Although this has to be the case at temperatures below Big Bang Nucleosynthesis (BBN), $T_{\rm BBN} \simeq \text{few} \; {\rm MeV}$~\cite{Kawasaki:2000en,Ichikawa:2005vw}, we have no direct information for the expansion rate and energy density make-up of the universe at higher temperatures. 

A motivated alternative history is an early matter dominated (MD) era, for example during inflationary reheating or moduli domination; The DM relic density in this case depends on two unknown temperatures: $T_M$, the temperature where the new form of matter takes over the energy budget, and $T_R$, the ``reheat temperature'' below which we recover the standard history (for details see e.g. Refs.~\cite{Co:2015pka,Co:2016vsi}). The resulting effect is a depletion of the DM relic density due to the entropy dumped in the plasma~\cite{McDonald:1989jd,Kamionkowski:1990ni,Chung:1998rq,Giudice:2000ex,Acharya:2009zt,Co:2015pka,Hamdan:2017psw}. As a consequence, for an early MD epoch the observed DM abundance is reproduced for {\it smaller} interaction rates between DM and plasma particles, with a consequent reduction of all detection signal strengths in DM searches.

The opposite conclusion applies to alternate cosmological histories where the universe expands very fast, as in presence of a new species $\phi$ whose energy density red-shifts as $\rho_\phi \propto a^{-(4+n)}$ (with $a$ the scale factor and $n > 0$). If equality between the energy density associated with the species $\phi$ and with radiation happens below the freeze-out temperature, DM is produced during this new cosmological phase. The relic DM density from freeze-out is then significantly larger than the one obtained by a standard calculation~\cite{Salati:2002md,Profumo:2003hq,DEramo:2017gpl,Dutta:2017fcn}, as a result of equality between the (faster) expansion rate and the thermal processes rates occurring at earlier times (i.e. at higher temperatures, when the comoving DM density is larger). The key consequence for DM phenomenology is that that {\it larger} couplings (and therefore larger predicted experimental signals, for example for the annihilation rate of dark matter pairs in the late universe) are needed to produce the observed DM abundance. Moreover, Ref.~\cite{DEramo:2017gpl} identified a completely new phenomenon that happens for large enough $n$: unlike the standard case, DM particles {\it keep annihilating} even long after the departure from chemical equilibrium. This novel behavior was dubbed {\it relentless} DM, and it was later confirmed by Ref.~\cite{Dutta:2017fcn}. Relentless DM also generically features larger-than-usual DM interaction rates with ordinary particles~\cite{DEramo:2017gpl}.

The subject of this work is DM freeze-in, another motivated mechanism for DM genesis where the relic density, in a standard cosmological setting can be calculated directly from the DM particle physics properties~\cite{Hall:2009bx,Bernal:2017kxu}. The same caveat as above applies to this case: a modified, non-standard thermal history will affect the predicted final density of DM from freeze-in. Our goal here is to perform a general analysis of DM freeze-in in a fast expanding universe, similarly to what we performed for freeze-out in Ref.~\cite{DEramo:2017gpl}, and to draw the critical phenomenological consequences of such scenario. DM particles produced through freeze-in are very weakly coupled with the primordial plasma and never attain thermal equilibrium in the early universe. Although very weak, the interactions with bath particles $B_i$ are enough to create DM particles $\chi$ through reactions $B_i \rightarrow \chi$. After  $\chi$ is produced, it simply red-shifts away and it is still present today contributing to the observed DM energy density.

The set of cosmological histories considered in this work is phenomenologically described by the two-dimensional parameter space $(T_r, n)$. Here, $T_r$ is the temperature where the energy density of $\phi$ equals the one of the radiation bath, whereas $n>0$ is the index describing how  the fluid red-shifts through the relation $\rho_\phi \propto a^{-(4+n)}$. These two parameters cannot take arbitrary values, since they are bound by BBN constraints~\cite{DEramo:2017gpl} which constrains the Hubble expansion rate, and hence the energy density of the universe at temperatures around when BBN operates to be close to pure radiation-domination.

In the spirit of a very general analysis, we consider the following freeze-in scenarios to produce DM particles $\chi$ through reactions involving bath particles $B_i$:
\begin{enumerate}
\item \textbf{Decay $B_1 \rightarrow B_2 \chi$:} a bath particle $B_1$, heavier than $\chi$, decays to a final state involving one DM particle and other bath particles (which we indicate generically with the symbol $B_2$). While we consider a two-body decay for illustration, our results are valid for general $n$-body decays. The discussion for decay channels involving more than one DM particle (e.g. $B_1 \rightarrow \chi \chi$) in the final state is analogous.
\item \textbf{Single Production $B_1 B_2 \rightarrow B_3 \chi$:} Collisions between two bath particles lead to {\it one} DM particle in the final state. This reaction happens, for example, when one initial state bath particle shares the same discrete quantum number with $\chi$, e.g. when both $B_1$ and $\chi$ are odd under a $\ZZ_2$ symmetry. 
\item \textbf{Pair Production $B_1 B_2 \rightarrow \chi \chi$:} Collisions between two bath particles lead to {\it two} DM particles in the final state. We separate this case from the one above since it happens in different theories. As an example, $\chi$ can be the {\it only}  particle odd under a $\ZZ_2$ symmetry, and it thus needs to be pair produced.
\end{enumerate}

A consistent picture emerges from our analysis of different cosmological histories and of various freeze-in scenarios: the observed DM abundance is reproduced for  {\it larger} couplings between DM and plasma particles compared to standard cosmological histories. This conclusion was also reached for freeze-in in an early MD epoch~\cite{Monteux:2015qqa,Co:2016fln,Co:2017orl}. A comparison among different cases is provided in Tab.~\ref{tab:tableintro}. The general, key conclusion of our study is that {\it  DM genesis in a fast expanding universe, be it via freeze-out or via freeze-in, always requires larger couplings, with the inescapable prediction of enhanced signals for DM detection}. 
\begin{center}
\begin{table}
\begin{center}
\begin{tabular}{ c||c|c }
$g / g_{\rm standard}$ & early MD era & fast-expanding universe \\
\hline \hline
DM freeze-out & smaller & larger \\  
DM freeze-in & larger & larger \\ 
\end{tabular}
\end{center}
\caption{Comparison between couplings needed to produce the observed DM abundance in a standard versus modified cosmological setting, for the two cases an early MD era and of a fast expanding universe. We consider both DM freeze-out and freeze-in, and for each case we identify whether the required coupling to the plasma is smaller or larger than the standard case.}
\label{tab:tableintro}
\end{table}
\end{center}

We note that freeze-in through pair production of DM particles (case 3 above) but limited to the specific case $n=2$ (kination domination) was studied in Refs.~\cite{Redmond:2017tja,Visinelli:2017qga}. The goal of this paper is to present, instead, a {\it general} analysis for different cosmological histories and freeze-in scenarios. For the particular case of $n=2$ our results are consistent with those presented in Refs.~\cite{Redmond:2017tja,Visinelli:2017qga}.

The reminder of this study has the following outline: After reviewing the Boltzmann equation for freeze-in with the modified cosmological background in Sec.~\ref{sec:BEFI}, we consider freeze-in production of DM in the early universe. As explicitly stated, we only focus on IR production (\textit{i.e.} production dominated by processes occurring at low temperatures, close to the bath particle masses). While this is always the case for decays, we identify under which circumstances IR production occurs from scattering as well. By focusing on IR production, we avoid issues related to the uncertain history of the universe {\it before} the time of $\phi$-domination. We then divide the following discussion into two parts: we deal with decay  in Sec.~\ref{sec:decays} and with scattering in Sec.~\ref{sec:scattering}. Wherever relevant, we highlight the  most prominent possible experimental signals associated with freeze-in within a non-standard cosmological history with faster-than-usual expansion rates at early times. We summarize our results in Sec.~\ref{sec:conclusions}. 

\section{Boltzmann Equation for Freeze-In}
\label{sec:BEFI}

The number density of DM particles $\chi$ evolves in an isotropic and homogeneous early universe according to the Boltzmann equation
\be
\frac{d n_\chi}{d t} + 3 H n_\chi = \mathcal{C}_\alpha  \ .
\label{eq:BEforFI}
\ee
The second term on the left-hand side accounts for the Hubble expansion, whereas number-changing reactions which, here, produce DM particles are accounted for by the collision operator on right-hand side. This collision operator $\mathcal{C}_\alpha$ depends on the specific reaction under consideration (e.g. $\alpha = B_1 \rightarrow B_2 \chi$). It also generically depends on time, or, equivalently, on the temperature of the radiation bath. 

The boundary condition we assume for the Boltzmann equation (\ref{eq:BEforFI}) is a vanishing DM number density at very early times. In other words, we are assuming here that physics at high scale (e.g. inflation) produces a negligible number of $\chi$ particles, which are then exclusively produced in the later universe by the freeze-in reactions listed in the Introduction. 

It is convenient to re-cast the Boltzmann equation factoring out the effect of expansion. To this end, as customary, we define the {\it comoving number density} $Y_\chi = n_\chi / s$, where $s$ is the entropy density. Using the definition of the comoving density, together with the assumed conservation of entropy, $s a^3 = {\rm const}$, we rewrite the Boltzmann equation as
\be
\frac{d Y_\chi}{d \log T} = -\left( 1 +  \frac{1}{3} \dfrac{\partial \log g_{*s}}{\partial \log T} \right)  \frac{\mathcal{C}_\alpha}{H \, s}    \ .
\label{eq:BEforFI2}
\ee
Finally, we introduce the dimensionless ``time variable'' $x = m_B / T$, where $m_B$ is typically the mass scale of some bath particles that we will specify for each case. Upon using the general relation $d f / d \log T = - d f / d \log x$, we find the final form of the Boltzmann equation
\be
\frac{d Y_\chi}{d \log x} = 
\left( 1 - \frac{1}{3} \dfrac{\partial \log g_{*s}}{\partial \log x} \right) \frac{\mathcal{C}_\alpha(x)}{H(x) \, s(x)}  \ ,
\label{eq:BEforFI3}
\ee
where we make explicit the $x$-dependence (i.e. time, or inverse temperature) of the Hubble parameter $H$, the entropy density $s$ and the collision operator $\mathcal{C}_\alpha$. 

In the next Sections, we specify each time our choice for $x$ and what reaction $\alpha$ we are considering. Before  discussing the freeze-in process, we conclude this Section with a brief review of the cosmological background and  a comparison between IR and UV production.

\subsection{The cosmological background}

We are interested in DM production for cosmological histories where the universe is dominated by a new species $\phi$, whose red-shift behavior is $\rho_\phi \propto a^{-(4+n)}$. Since entropy is conserved, during the time of $\phi$-domination the energy density scales as $\rho_\phi \propto T^{-(4+n)}$, where $T$ is the temperature of the radiation bath. The Friedmann equation allows us to identify the relation $H \propto T^{-(2+n/2)}$. 

Motivated theories leading to this faster Hubble expansion can be found in models for dark energy and/or inflation (see e.g. Refs.~\cite{DeSantiago:2011qb,Ratra:1987rm,Wetterich:1987fm,Caldwell:1997ii,Sahni:1999gb,Choi:1999xn,Kamenshchik:2001cp,Khoury:2001wf,Gardner:2004in,Chavanis:2014lra,Dimopoulos:2017zvq,Dutta:2017fcn}. Famously, quintessence theories explaining the current acceleration feature an early phase where the universe is dominated by the kinetic energy of a new scalar field (kination regime), which is equivalent to the case $n = 2$ in our parameterization. An even faster expansion can be achieved in the context of ekpyrotic scenarios, since larger values $n>2$ are needed to smooth the universe out in the contracting phase. An explicit example of a microscopic theory leading to the $n>2$ expansion was provided in Ref.~\cite{DEramo:2017gpl}.

Regardless of the microscopic theory,  the Hubble parameter at a fixed temperature is always larger than its associated value for a standard history at the same temperature when the universe is dominated by $\phi$. This is why the cosmological histories considered in this work are the ones for a {\it fast expanding universe}. A complete description of these histories and how BBN bounds the parameter space can be found in Ref.~\cite{DEramo:2017gpl}. Here, we summarize the key results.

When the universe is dominated by $\phi$, the Hubble parameter at a fixed temperature is always larger than its associated value for a standard history at the same temperature. This is why the cosmological histories considered in this work are the ones for a {\it fast expanding universe}. A complete description of these histories and how BBN bounds the parameter space can be found in Ref.~\cite{DEramo:2017gpl}. Here, we summarize the key results.

The cosmological background is identified by two parameters: $(T_r, n)$. The temperature $T_r$ is set by some boundary condition, and we choose it to be the temperature where the energy density of $\phi$ and radiation are the same. The index $n$ described the red-shift behavior. The energy density of $\phi$ as a function of the radiation bath temperature is given by
\be
\rho_\phi(T) = \rho_\phi(T_r)  \left(\frac{g_{*s}(T)}{g_{*s}(T_r)}\right)^{(4+n)/3} \left(\frac{T}{T_r}\right)^{(4+n)} \ .
\ee
The total energy density at any temperature reads
\be
\begin{split}
& \rho(T) = \rho_{\rm rad}(T) + \rho_\phi(T) = \\ & 
\rho_{\rm rad}(T) \left[1 + \frac{g_*(T_r)}{g_*(T)} \left( \frac{g_{*s}(T)}{g_{*s}(T_r)} \right)^{(4+n)/3} \left(\frac{T}{T_r}\right)^{n} \right] \ ,
\label{eq:rhototal}
\end{split}
\ee 
where we factor out the energy density of the radiation bath. The Hubble parameter as a function of the temperature can be computed using Friedmann's equation
\be
H = \frac{\sqrt{\rho}}{\sqrt{3} \, M_{\rm Pl}}  \ ,
\label{eq:HubbleRate}
\ee
where the reduced Planck mass is $M_{\rm Pl} = (8 \pi G)^{-1/2} = 2.4 \times 10^{18} \, {\rm GeV}$. This is the expression for the Hubble parameter, with the energy density $\rho$ as given in \Eq{eq:rhototal}. The Hubble parameter $H(T)$ enters the Boltzmann equation~\eqref{eq:BEforFI3}, which we use to compute the DM relic density. All results in this paper are obtained via a numerical calculation with this complete expression for the Hubble parameter. However, in order to perform simple analytical estimate, it is useful to give an approximate expression for the Hubble rate at temperatures larger than $T_r$
\beq
H(T) \simeq \frac{\pi \, \overline{g}^{1/2}_*}{3 \sqrt{10}} \dfrac{T^2}{M_{\rm Pl}} \left(\frac{T}{T_r}\right)^{n/2} \ ,  \qquad
(T \gg T_r) \ ,
\label{eq:HubbleEarlyTimes}
\eeq
where we take $g_{*s}(T) = g_*(T) = \overline{g}_* = {\rm const}$. The full matter content of the Standard Model gives $\overline{g}_* = g_{*{\rm SM}} =106.75$. Finally, as found in Ref.~\cite{DEramo:2017gpl}, BBN bounds the cosmological parameters to be
\be
T_r \gtrsim (15.4)^{1/n} \; {\rm MeV} \ .
\label{eq:BBNbound}
\ee

\begin{figure}
\center\includegraphics[width=0.483 \textwidth]{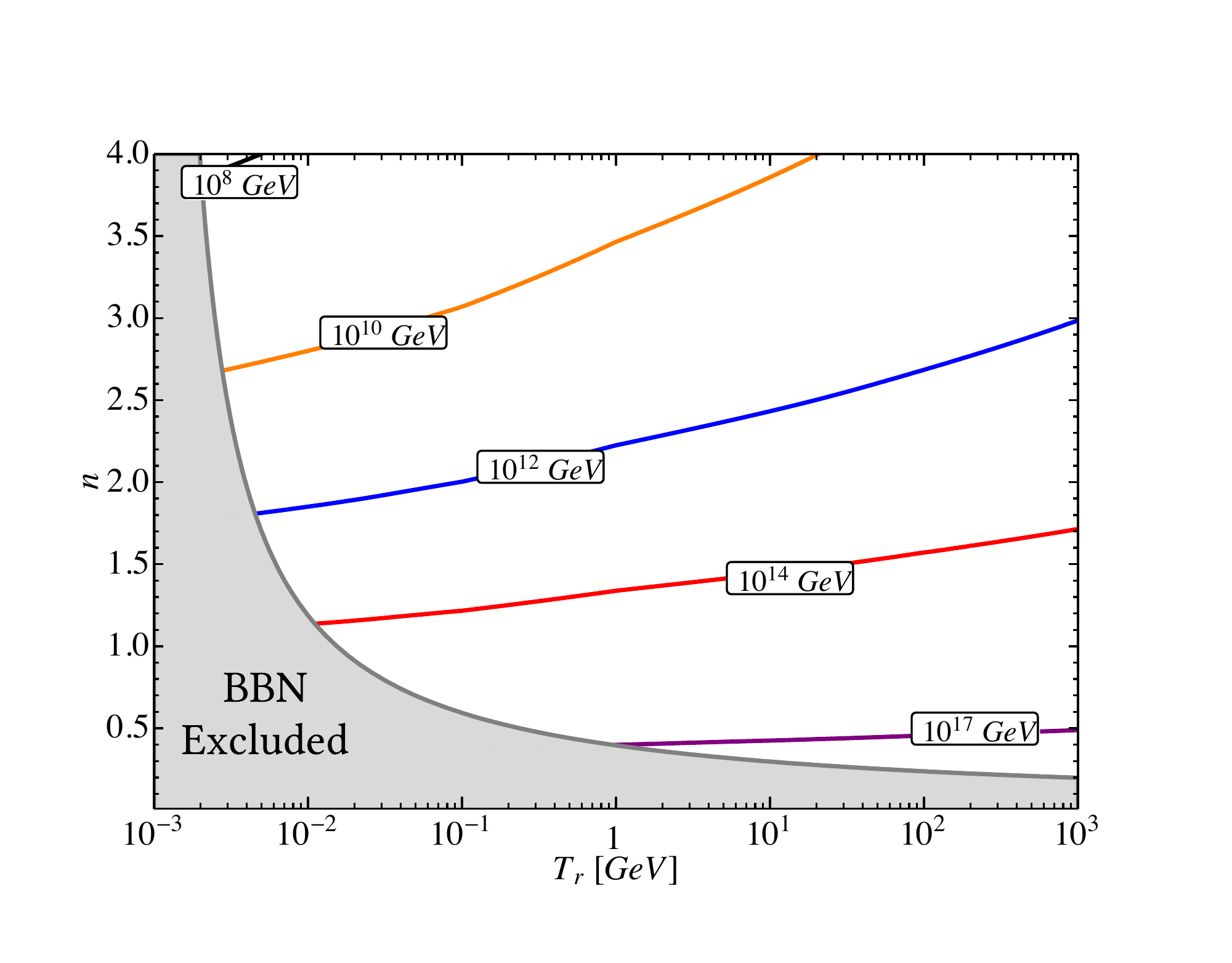}
\caption{Upper bound on the reheat temperature after inflation $T_{\rm RH}$ in the $(T_r, n)$ plane. The gray region is excluded by BBN.}
\label{fig:Reheating}
\end{figure}

The cosmological history introduce here and parameterized by $(T_r, n)$ cannot be extrapolated arbitrarily back in time. As we consider a younger universe, or equivalently as we go to higher temperature, the energy density of $\phi$ gets larger. We identify the temperature when we reach the value $\rho_\phi \sim M_{\rm Pl}^4$, and we set this as a limit above which we cannot use our framework anymore. This in turns imply a constraint on $T_{\rm RH}$, the reheating temperature after inflation. We find the bound 
\be
T_{\rm RH} \lesssim  M_{\rm Pl}  \, (T_r/M_{\rm Pl}  )^{n/(n+4)} \ .
\ee
In Fig.~\ref{fig:Reheating}, we visualize this upper bound on the reheating temperature in the $(T_r, n)$ plane. We consider the region of this plane correspondent to the set of cosmological histories analyzed in the following study, and we also shade away the region excluded by the BBN bound in \Eq{eq:BBNbound}. In all the parameter space of interest, this bound is several orders of magnitude above the masses of the particles considered in this work. As explained in the following sub-section, we will only consider IR production, namely freeze-in processes mostly active at temperatures around the typical masses of the particles involved in the reactions. Thus we can safely assume that $T_{\rm RH}$ is well above the masses under consideration, but still well below the upper bound given in Fig.~\ref{fig:Reheating}.

\subsection{IR vs. UV production}

A remarkable feature of freeze-in is that DM production, with a standard cosmological history, is always IR dominated~\cite{Hall:2009bx}. In this section we show that this is always the case for freeze-in from decays, even in the case of a modified cosmological history with a fast-expanding universe at early times. If DM particles are produced via scattering processes, instead, the production with a standard cosmological history is IR dominated as long as the interactions between DM and the bath particles are renormalizable. We conclude this Section with a comparison between IR and UV production for the cosmological histories considered in this work. 

Our assumption through this work is that at very high temperatures the abundance of $\chi$ is negligible. As the temperature drops down, DM particles are created via processes involving the plasma particles. At a given temperature $T$, much higher than the DM and the bath particles masses, the comoving abundance of $\chi$ particles approximately reads
\be
Y_\chi(T) \simeq \gamma(T) \, H(T)^{-1} \ .
\ee
Here, $\gamma(T)$ is the (temperature dependent) rate for the process under consideration, whereas the inverse Hubble parameter is about the age of the universe. This simple relation allows to establish whether the production is IR or UV dominated. 

We start from the case of decays, $B_1 \rightarrow B_2 \chi$, where the rate scales as $\gamma_{B_1 \rightarrow B_2 \chi}(T) \simeq \Gamma_{B_1 \rightarrow B_2 \chi} \, m_{B_1} / T$. The partial width computed in the rest frame of $B_1$ is corrected by the Lorentz time dilatation factor. Upon using the approximate Hubble parameter in \Eq{eq:BBNbound}, and neglecting numerical factors, we find for decays
\be
\left. Y_\chi(T)\right|_{B_1 \rightarrow B_2 \chi} \simeq  \Gamma_{B_1 \rightarrow B_2 \chi} \, 
\frac{m_{B_1} M_{\rm Pl}\, T_r^{n/2} }{T^{3+n/2}} \ .  
\ee
Thus freeze-in from decays is always dominated at low (IR) temperatures.

For the case of scattering, the temperature dependence of the rate stems from the type of interaction under consideration. If we take an operator of mass dimension $d$ as responsible for the scattering process, the rate scales $\gamma_{B_1 \rightarrow B_2 \chi}(T) \propto T^{2d - 7} / M_*^{2d - 8}$, where $M_*$ is the mass scale appearing in the operator. The comoving density scales with the temperature as
\be
\left. Y_\chi(T)\right|_{B_1 B_2 \rightarrow B_3 \chi} \propto \frac{T^{2d - 9 - n/2}}{M_*^{2d - 8}} \frac{M_{\rm Pl}}{T^{n/2}_r} \ .
\ee
The scaling for the case $B_1 B_2 \rightarrow \chi \chi$ is identical. Thus for freeze-in via scattering the production is IR dominated only for operators whose mass dimension satisfies
\be
d < 4.5 + \frac{n}{4} \ .
\label{eq:ineq}
\ee
The case $n=0$ corresponds to a standard history, and for this case freeze-in is IR dominated only for renormalizable interactions, as correctly identified in Ref.~\cite{Hall:2009bx}.

We always consider IR production in this work. And in doing so we avoid the complication of specifying how the cosmological phase of $\phi$ domination arises at very high temperatures. All we assume here is that at temperatures above the plasma particle masses $\phi$ domination sets in, and DM particles are produced at around the mass scale of the bath particles. As discussed above, this is automatic for decays, whereas for scattering IR production only applies for interactions satisfying \Eq{eq:ineq}. The comoving number density at any ``time'' $x$ can be computed from \Eq{eq:BEforFI4} by solving a numerical integral
\be
Y_\chi(x) = \int_0^x \frac{dx^\prime}{x^\prime}\left( 1 - \frac{1}{3} \dfrac{\partial \log g_{*s}}{\partial \log x^\prime} \right)  \frac{\mathcal{C}_\alpha(x^\prime)}{H(x^\prime) \, s(x^\prime)}   \ .
\label{eq:BEforFI4}
\ee
Here, the lower integration extreme ($x^\prime = 0$) is justified by IR production. The final DM density is given by taking $x \rightarrow \infty$ in the above equation.

\section{Freeze-In from Decays}
\label{sec:decays}

We start with the case where DM particles are produced through the decay process
\be
B_1 \rightarrow B_2 \chi \ .
\label{eq:decayreaction}
\ee
We provide a complete derivation of the collision operator for this process in \Eq{eq:appdecay} of App~\ref{app:thermalaverages}. Here, we only quote the final result,
\be
\mathcal{C}_{B_1 \rightarrow B_2 \chi} = n_{B_1}^{\rm eq} \, 
\Gamma_{B_1 \rightarrow B_2 \chi}  \, \frac{K_1[m_{B_1}/T]}{K_2[m_{B_1}/T]}  \ . 
\ee
For the case of decays, it is convenient to choose $x = m_{B_1} / T$. Furthermore, we take the equilibrium distribution from \Eq{eq:appneq}, and we rewrite the collision operator for decays as a function of the variable $x$
\be
\mathcal{C}_{B_1 \rightarrow B_2 \chi} = \frac{g_{B_1} \, m_{B_1}^3}{2 \pi^2} \, \frac{K_1[x]}{x} \, 
\Gamma_{B_1 \rightarrow B_2 \chi}  \ . 
\ee

The comoving density at any temperature can be computed by applying the general result in \Eq{eq:BEforFI4}. After plugging the explicit expression for the entropy density, the freeze-in comoving density reads
\be
\begin{split}
Y_\chi(x) = & \, g_{B_1} \, \frac{45}{4 \pi^4} \,  \Gamma_{B_1 \rightarrow B_2 \chi}  \times \\ &
\int_0^x dx^\prime \left( 1 - \frac{1}{3} \dfrac{\partial \log g_{*s}}{\partial \log x^\prime} \right)   \frac{K_1[x^\prime] \, x^{\prime}}{g_{*s}(x^\prime) \, H(x^\prime)} \ .
\label{eq:Ydecay}
\end{split}
\ee
This is our master equation to compute freeze-in production via decays. The only assumption so far is that entropy is conserved, thus this equation is also valid for the case of a standard thermal history. The details of the thermal history under consideration enter through the Hubble parameter $H(x^\prime)$ in the denominator of the integrand. 

\subsection{Number Density Evolution}

We parameterize the partial decay width with the expression
\be
\Gamma_{B_1 \rightarrow B_2 \chi}  = \frac{\lambda_d^2}{8 \pi} m_{B_1} \ .
\label{eq:decaywidth}
\ee
Here, $\lambda_d \ll 1$ is a very small coupling mediating the decay process, whereas the factor of $8\pi$ in the denominator accounts for the phase space of the two-body final state. 

\begin{figure}[t]
\center\includegraphics[width=0.483 \textwidth]{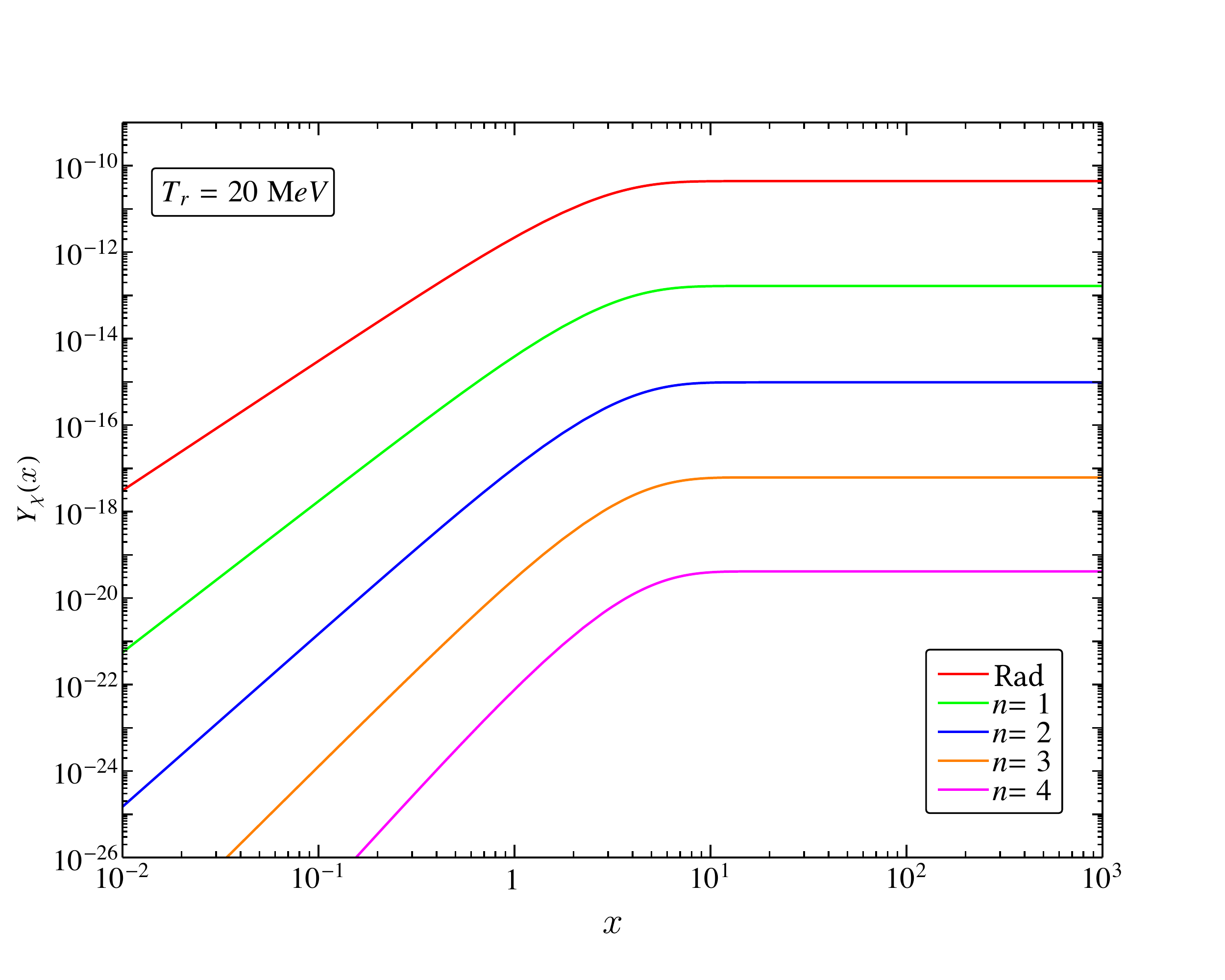}
\caption{Numerical solutions for the comoving number density $Y_\chi$ for the case of freeze-in from decays. We choose $g_{B_1} = 2$, $m_{\chi}=10\gev$, $m_{B_1}=1\tev$, and $\lambda_{d} = \lambda_{d}^{\rm rad} = 1.22 \times 10^{-11}$. We always set $T_{r}=20 \mev$.}
\label{fig:FI_Decay}
\end{figure}

The freeze-in number density of $\chi$ particles is determined by \Eq{eq:Ydecay} once we specify the mass of the decaying particle and the coupling $\lambda_d$. The asymptotic value for the number density is found by taking the $x \rightarrow \infty$ limit, whereas the energy density is obtained by just multiplying the previous result by the mass of $\chi$. As an illustrative example, we fix $(m_{B_1}, m_\chi) = (1000, 10) \, \gev$, and we also fix $g_{B_1} = 2$. The observed DM abundance for a standard cosmological history is achieved if we choose $\lambda^{\rm rad}_{d}=1.22 \times 10^{-11}$. In Fig.~\ref{fig:FI_Decay} we keep these particle physics parameters constant, and we show numerical solutions for different modified cosmological histories. We always take $T_{r} = 20 \mev$, consistently with BBN bounds, and we show solutions for different values of $n$. 

Fig.~\ref{fig:FI_Decay} illustrates well our findings: the asymptotic comoving density consistently decreases as we increase the value of the index $n$.  In a fast expanding universe, the freeze-in production from bath particle decays is less effective than in the case of a standard cosmological history (red line). As a result, larger couplings are required to reproduce the observed DM density. Moreover, for each given temperature the comoving density is always lower as we consider larger values of $n$. Correspondingly, the same freeze-in yield is achieved at lower temperatures. 

The results in Fig.~\ref{fig:FI_Decay} are readily explained by an approximate solution to the Boltzmann equation. (We remind the Reader that what shown in the plot was obtained by numerically solving the integral in \Eq{eq:Ydecay}). It is helpful to recall the asymptotic behavior for the Bessel function appearing in the integrand
\be
K_1[x^\prime] \simeq \left\{\begin{array}{ccccc}
\frac{1}{x^\prime} & & & & x^\prime \ll 1 \\
\sqrt{\frac{\pi}{2 x^\prime}} e^{- x^\prime}  & & & & x^\prime \gg 1 \\
\end{array} \right.
\label{eq:BesselK1}
\ee
The physics behind the suppression at large $x^\prime$, namely at temperature much lower than the decaying particle mass, is clear: decaying particles are exponentially rare at temperatures below their mass, thus freeze-in production in this range of temperatures is negligible. As a result, in Fig.~\ref{fig:FI_Decay} the comoving yields are just horizontal lines at $x^\prime \gg 1$: the integral in \Eq{eq:Ydecay} is saturated around $x^\prime \simeq 1$. 

For all cases in Fig.~\ref{fig:FI_Decay} we also have $T_r \ll m_{B_1}$, thus freeze-in production happens entirely during the phase of $\phi$-domination. If we additionally neglect the temperature variation for the number of relativistic degrees of freedom, namely we set $g_{*s}(x) = g_*(x) = \overline{g}_*$, we can rewrite \Eq{eq:Ydecay} as follows
\be
\begin{split}
Y_\chi(x) \simeq & \, \frac{g_{B_1}}{\overline{g}^{3/2}_*} \, \frac{135 \sqrt{10}}{4 \pi^5} \, \frac{\Gamma_{B_1 \rightarrow B_2 \chi} \, M_{\rm Pl}}{m_{B_1}^2 x_r^{n/2}}  \times \\ &
\int_0^x dx^\prime K_1[x^\prime] \, x^{\prime \,(3 + n/2)} \ ,
\end{split}
\label{eq:YFIdecayapprox}
\ee
where we introduce $x_r = m_{B_1} / T_r$. The asymptotic value for the comoving density can be computed analytically. We write it as follows:
\be
Y_\chi^\infty  = \left. Y_\chi^\infty\right|_{\rm rad} \times  \mathcal{F}_{\rm decay}(T_r, n) \ , 
\label{eq:Ychidecayanalytical}
\ee
where we calculate the suppression factor $\mathcal{F}_{\rm decay}$ with respect to the result in a pure radiation dominated early universe~\cite{Hall:2009bx}
\be
\left. Y_\chi^\infty\right|_{\rm rad} = \frac{g_{B_1}}{\overline{g}^{3/2}_*} \, \frac{405 \sqrt{10}}{8 \pi^4} \,  \frac{\Gamma_{B_1 \rightarrow B_2 \chi} \, M_{\rm Pl}}{m_{B_1}^2}  \ ,
\ee
and we define the function accounting for the correction
\be
\mathcal{F}_{\rm decay}(T_r, n) \equiv \frac{8}{3 \pi}\left(\frac{2}{x_r}\right)^{n/2} \varGamma\left[\frac{6 + n}{4}\right]  \varGamma\left[\frac{10 + n}{4}\right]  \ .
\label{eq:Fdecaydef}
\ee

Here, $\varGamma[x]$ is the Euler gamma function. This result is valid only for $n > 0$. Notice that we do not recover the radiation case result for $n=0$: this is consistent with the expression for the energy density in \Eq{eq:rhototal} where setting $n=0$ does not get rid of $\phi$, but, rather, it adds a new species that red-shifts like radiation. From the explicit expression for $\mathcal{F}(n)$ we immediately see that the main source for the difference among the horizontal lines location in Fig.~\ref{fig:FI_Decay} is the factor $x_r^{n/2}$ in the denominator, since for the case we consider we have $x_r = 5 \times 10^4$.

\begin{figure}[t]
\center\includegraphics[width=0.483 \textwidth]{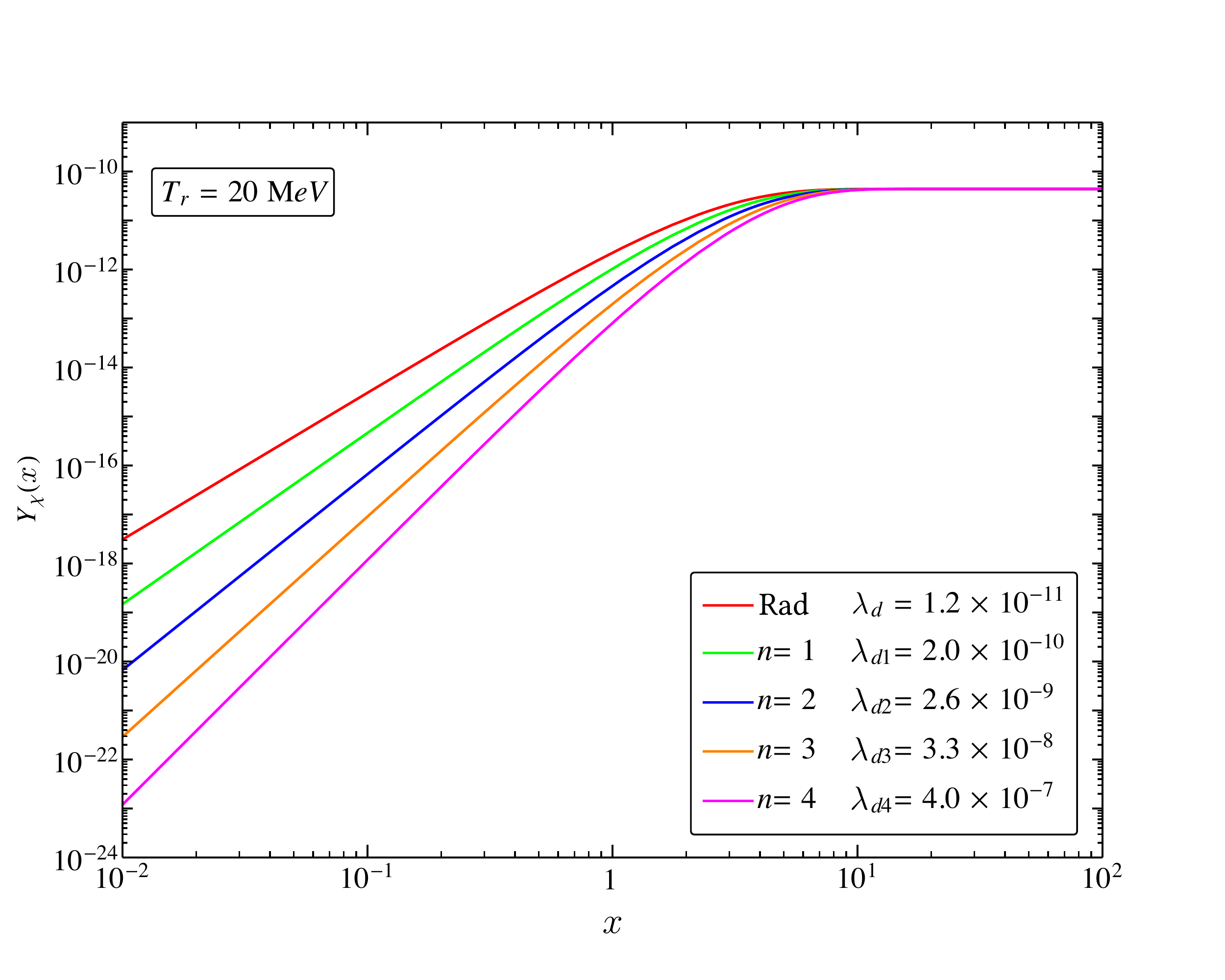}
\caption{Numerical solution for the comoving number density $Y_\chi$ with $m_{\chi}=10\gev$, $m_{B_1}=1\tev$. Now $\lambda_{d}$ is changed in order to reproduce the observed abundance ($\lambda_{d1}=2.0 \times 10^{-10}$, $\lambda_{d2}=2.6 \times 10^{-9}$, $\lambda_{d3}=3.3 \times 10^{-8}$, $\lambda_{d4}=4.0 \times 10^{-7}$).  We set $T_{r}=20 \mev$ for all $n$.}
\label{fig:FI_Decay_samerelic}
\end{figure}

The slope of the numerical solutions at $x \lesssim 1$ can also be derived analytically by taking the appropriate limit for the Bessel function (see \Eq{eq:BesselK1}). We consider \Eq{eq:YFIdecayapprox} in the $x \ll 1$ regime, where the integral is straightforward and we find 
\beq
Y_\chi(x) \propto x^{(3 + n/2)} \qquad \qquad (x \ll 1) \ .
\label{eq:FIscaling}
\eeq
The freeze-in solutions are steeper for larger $n$. The predicted asymptotic behavior is indeed what we find with our full numerical treatment in Fig.~\ref{fig:FI_Decay_samerelic}, where we take the same mass values for $B_1$ and $X$ but this time we choose the coupling $\lambda_d$ to reproduce the observed DM density. The steepness of the lines with larger $n$ allows freeze-in process to start later and to be dominated at slightly lower temperatures. 

\subsection{Relic Density Suppression}

\begin{figure}[H]
\center\includegraphics[width=0.483 \textwidth]{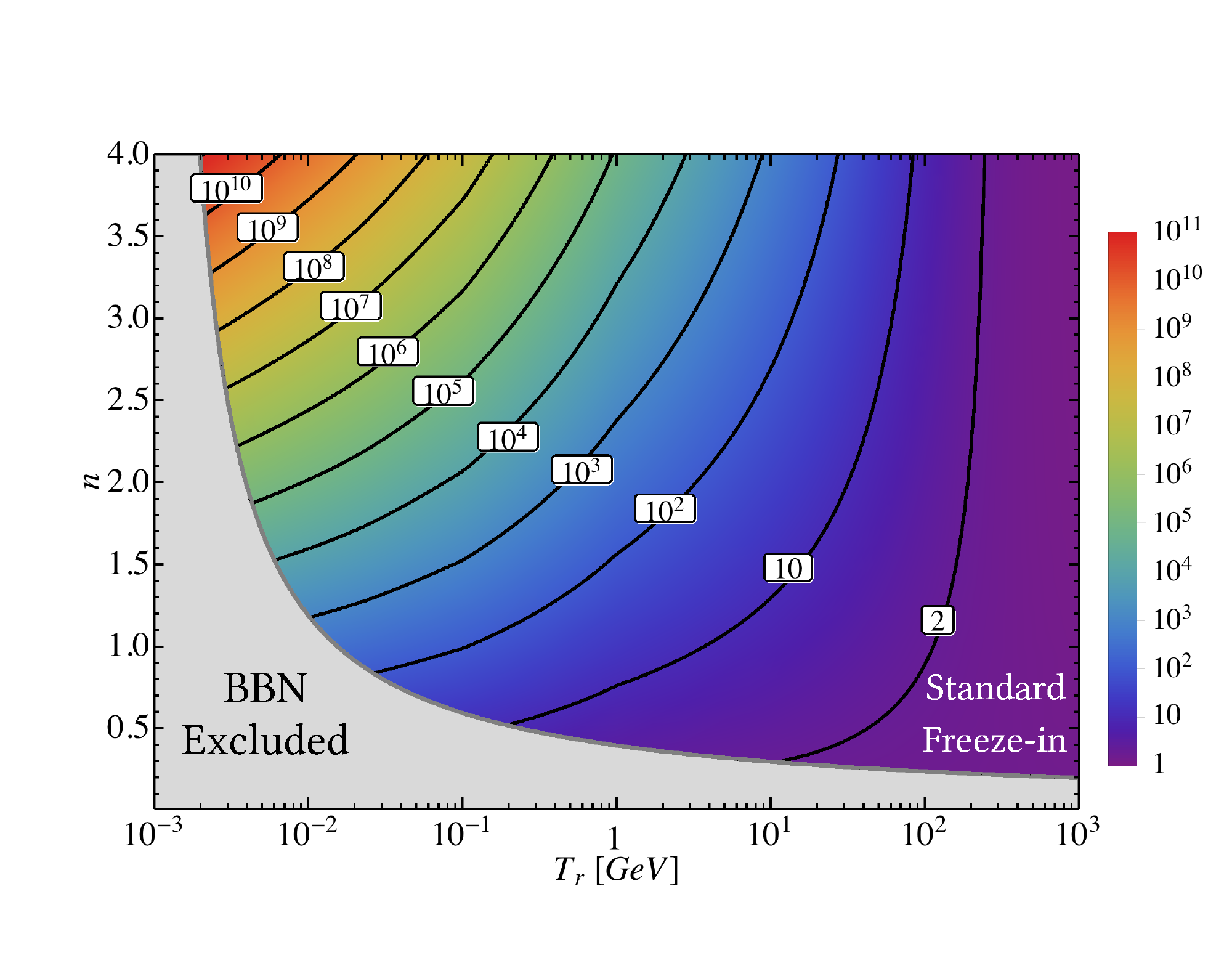}
\caption{Contour plots of the reduction in the relic density in the case of Freeze-in by decay.}
\label{fig:FI_Decay_E}
\end{figure}

Within the modified cosmological setup we consider in this work, DM is always under-produced with respect to the case of a standard history. We quantify by how much the relic density is suppressed in Fig. \ref{fig:FI_Decay_E}, where we keep the particle physics parameters fixed to the same values we used in the previous section. We calculate the DM relic density for each point in the $(T_r, n)$ plane, and we take the ratio between the observed DM relic density in the radiation case and the relic density in our modified cosmological setup. In other words, we show iso-countours for the function
\beq
r_{\rm decay}(T_r, n) \equiv \dfrac{\left.\Omega_\chi h^2\right|_{\rm rad}}{\Omega_\chi h^2}  \ . 
\eeq
For $T_r$ as large as $m_{B_1}$, the effect of the fast expanding universe phase is less important and we are back to a ``standard'' freeze-in scenario. For lower values of $T_r$, but still consistent with the BBN bound in \Eq{eq:BBNbound}, the factor can be as large as $10^{10}$. For these low values of $T_r$ we can approximate the result by using the semi-analytical solution found above
\be
r_{\rm decay}(T_r, n) \simeq \mathcal{F}_{\rm decay}(T_r, n)^{-1}  \qquad   \qquad (T_r \ll m_{B_1}) \ .
\ee

One could turn the argument around, and state that stronger interactions are needed to reproduce the observed DM density. The enhancement of the dimensionless coupling $\lambda_d$ defined in \Eq{eq:decaywidth} is easily obtained from Fig.~\ref{fig:FI_Decay_E}, since the final relic density is always proportional to the decay width. We find the relation
\beq
\lambda_d(T_r, n) =  r_{\rm decay}(T_r, n)^{1/2} \, \lambda_d^{\rm rad} \, ,
\eeq
with $\lambda_d^{\rm rad}$ the coupling for the case of a standard history. This enhancement to the couplings required to produce the right DM density today can thus be as large as $10^5$ with modified fast-expanding thermal histories.

\subsection{Displaced Events at Colliders}

We conclude this Section by commenting on the consequences of the coupling constant enhancement required for successful freeze-in DM production in modified cosmological settings. Once we fix the mass of the particles, the requirement of reproducing the observed relic density fixes the decay width for each point in the $(T_r, n)$ plane. The inverse decay width gives the scale for the decay length $\tau_{B_1} = \Gamma_{B_1}^{-1}$ if $B_1$ particles are produced at colliders. As we will see shortly, a typical prediction in the $(T_r, n)$ plane is the observation of displaced $B_1$ decay vertices at particle colliders. This is opposed to the case of a standard cosmology, where the decay width is too large and for collider purposes $B_1$ is a stable particle~\cite{Hall:2009bx}. Displaced events at collider are also typical is DM is produced via freeze-in during an early matter dominated era~\cite{Co:2015pka}.

A convenient variable to express the observed DM density is the comoving energy density
\be
\xi_\chi^{\rm obs} = \frac{m_\chi n_\chi}{s_0} = m_\chi Y_\chi = 0.44 \, {\rm eV} \ ,
\ee
with $s_0$ the current entropy density. We can find an approximate expression for the expected decay length by taking the solution in \Eq{eq:Ychidecayanalytical} and compare it with the value above
\be
\tau_{B_1} \simeq 3.4 \times 10^7 \, \mathcal{F}(T_r, n) \left(\frac{m_\chi}{10 \, {\rm GeV}}\right) \left(\frac{1 \, {\rm TeV}}{m_{B_1}}\right)^2 \, {\rm cm} \ ,
\label{eq:tauB1approx}
\ee
where we also fixed $\overline{g}_* = 106.75$ (accounting for the full SM degrees of freedom). The scale $10^7 \, {\rm cm}$, way above the size of any detector, is typical for freeze-in during a radiation dominated era. However, as observed above, for the cosmologies we consider in this work we typically have $\mathcal{F}(T_r, n) \ll 1$, thus we can potentially get back to the detector size. We actually know how much we can reduce this decay length, since the inverse of $\mathcal{F}(T_r, n)$ is what is shown in Fig.~\ref{fig:FI_Decay_E}. This suppression can be as large as $10^{10}$ and the decay length can get as small as $10^{-3} \, {\rm cm}$.

\begin{figure*}[htp]
\centering 

\hspace{-0.4cm}
\includegraphics[width=.282\textwidth]{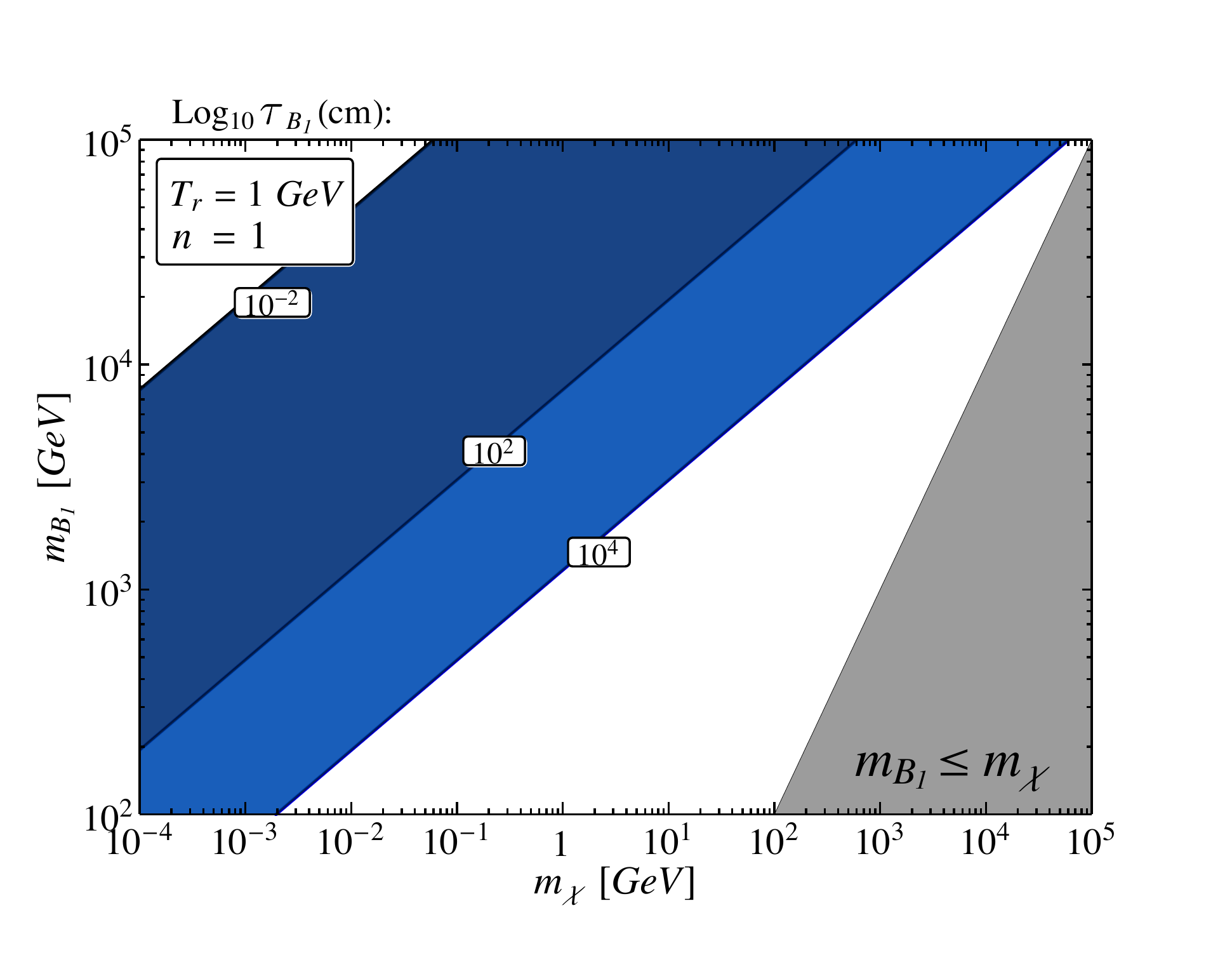}\hfill \hspace{-1.2cm}
\includegraphics[width=.282\textwidth]{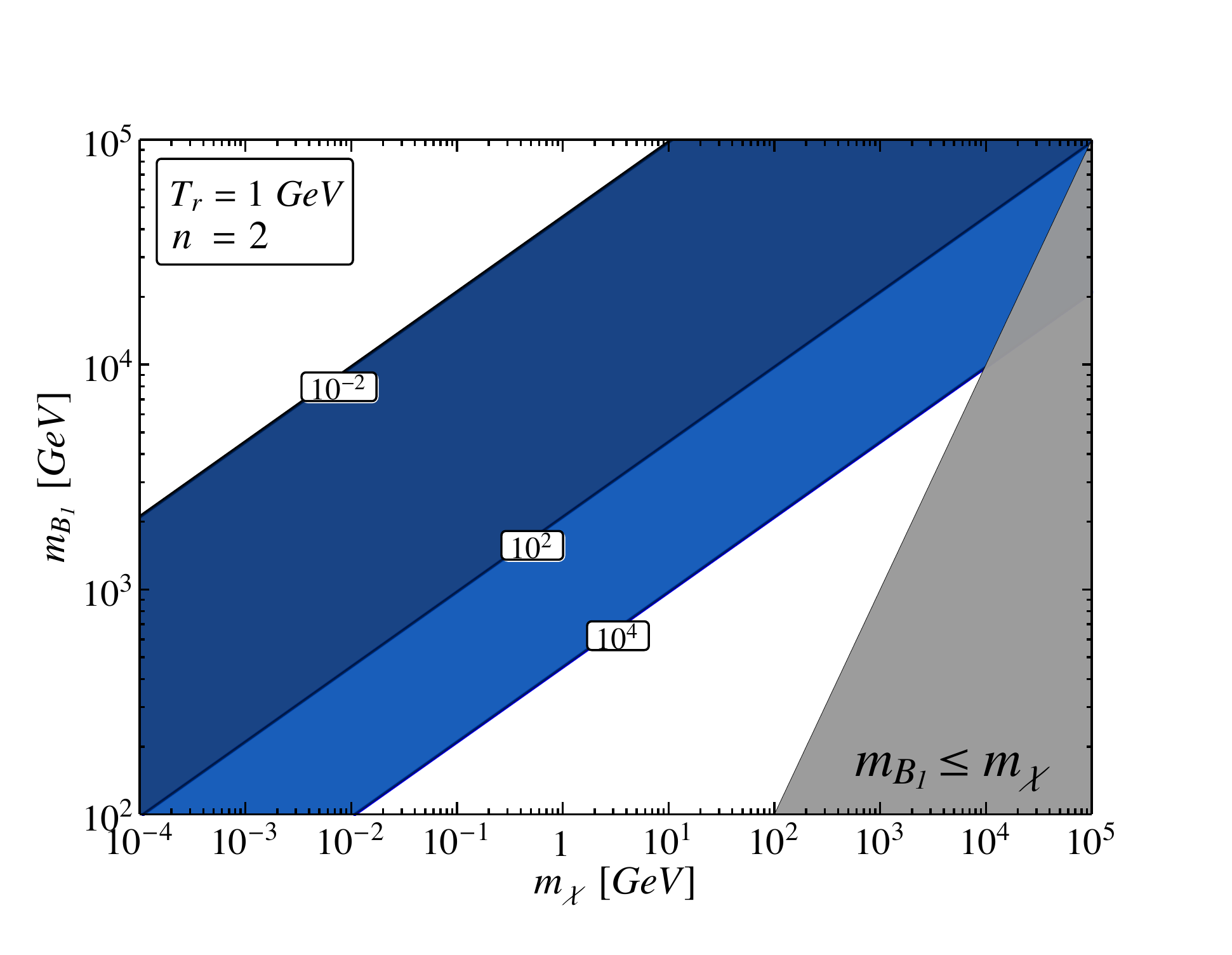}\hfill \hspace{-1.2cm}
\includegraphics[width=.282\textwidth]{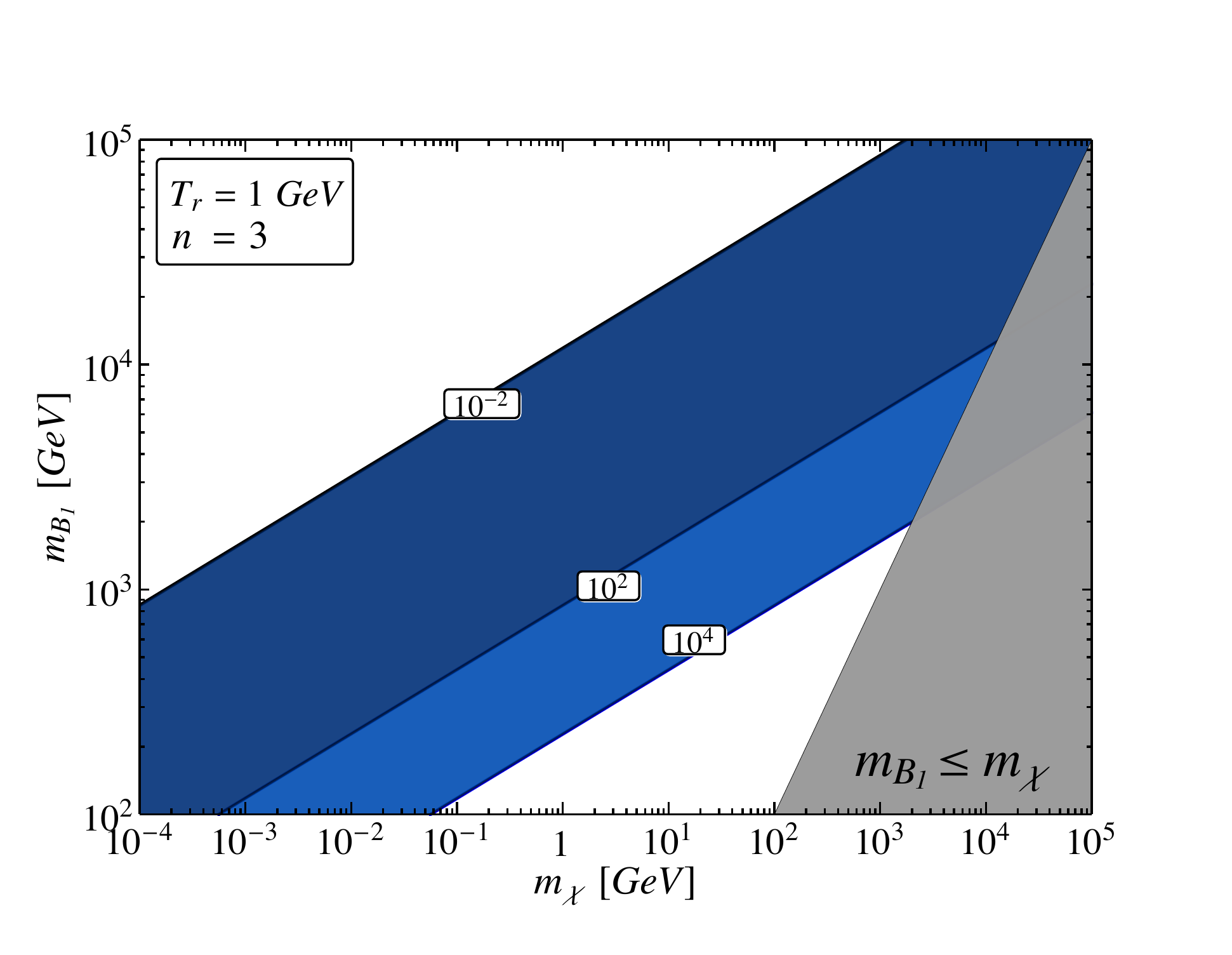}\hfill \hspace{-1.2cm}
\includegraphics[width=.282\textwidth]{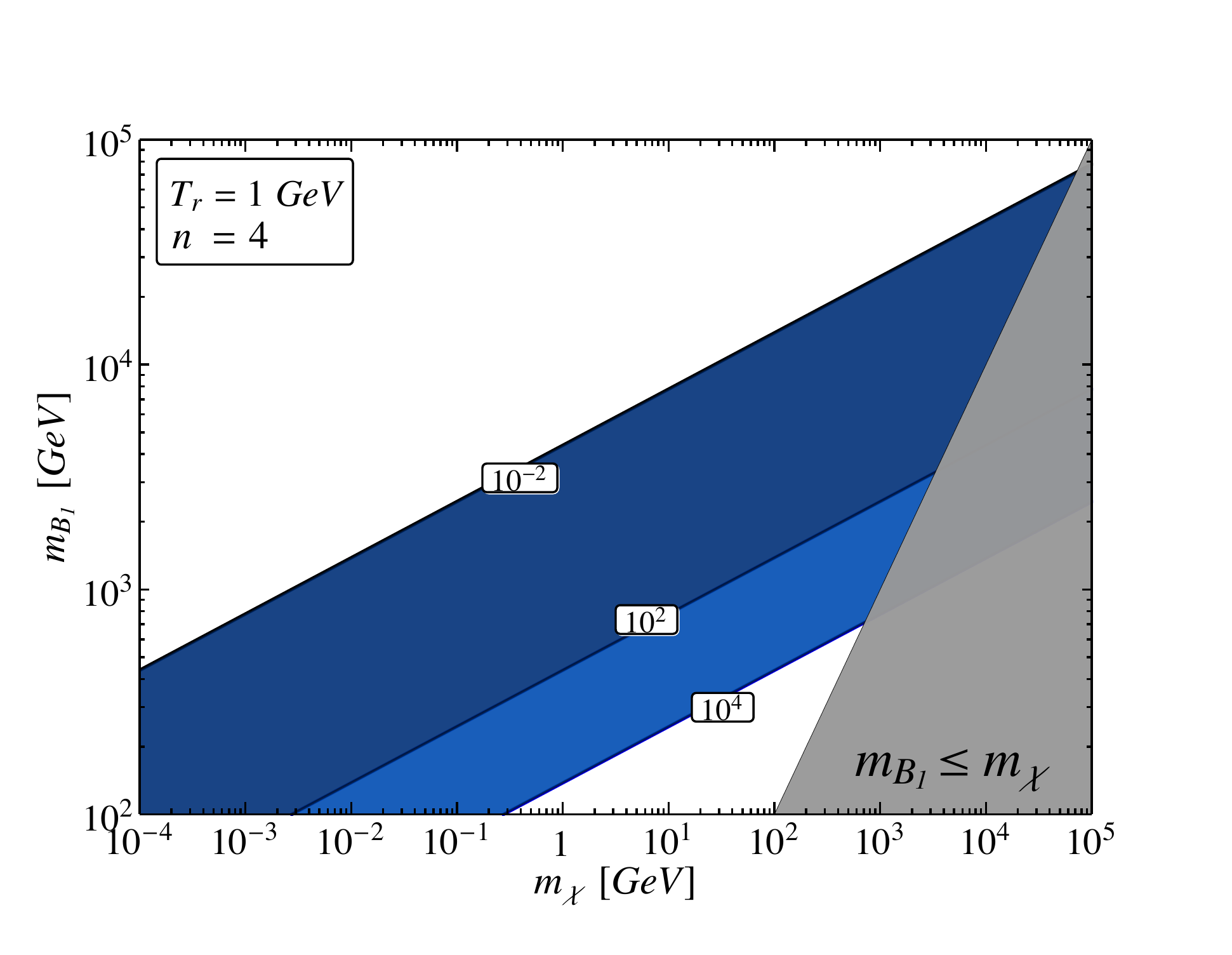}\hspace{-0.45cm}

\medskip

\hspace{-0.4cm}
\includegraphics[width=.282\textwidth]{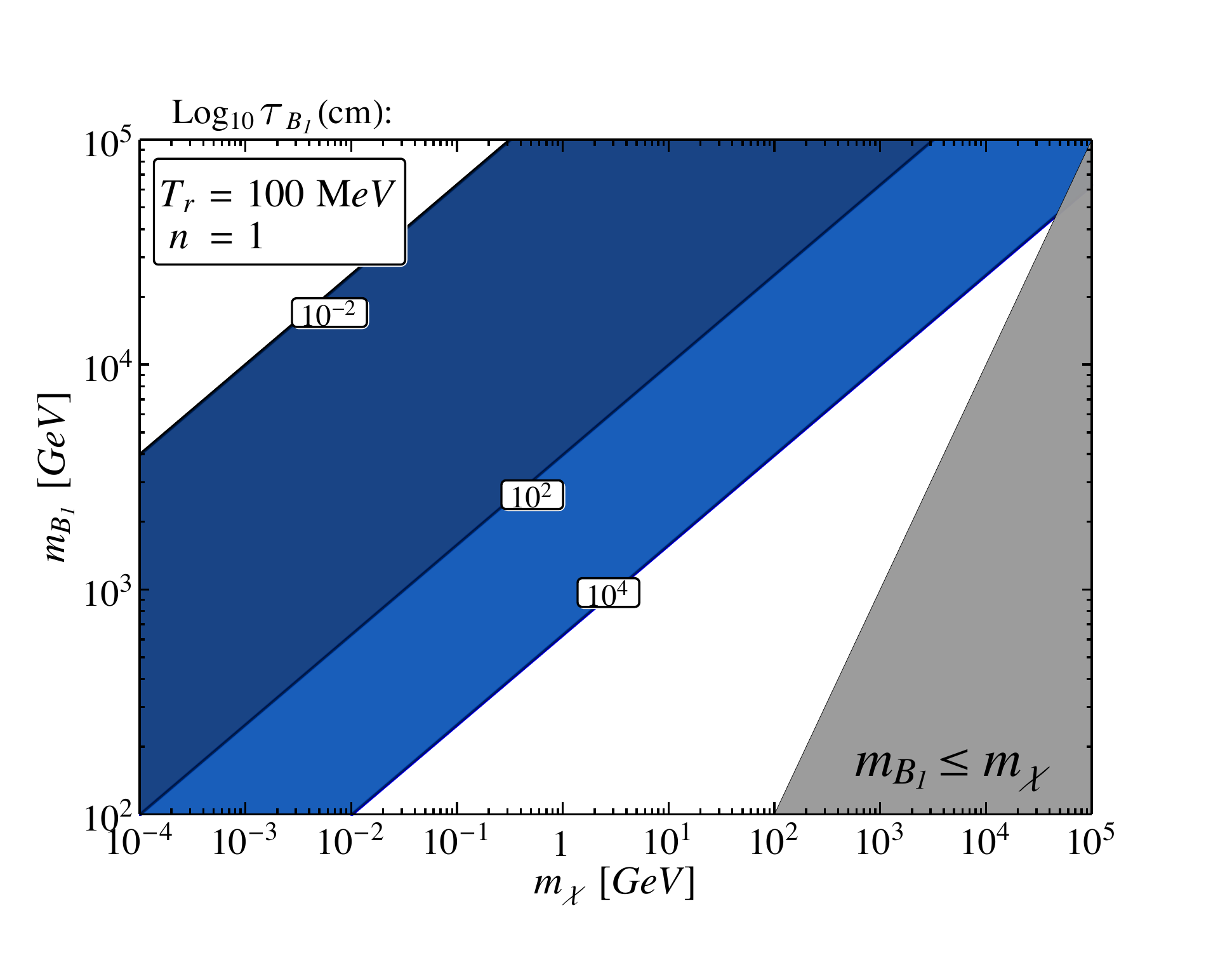}\hfill \hspace{-1.2cm}
\includegraphics[width=.282\textwidth]{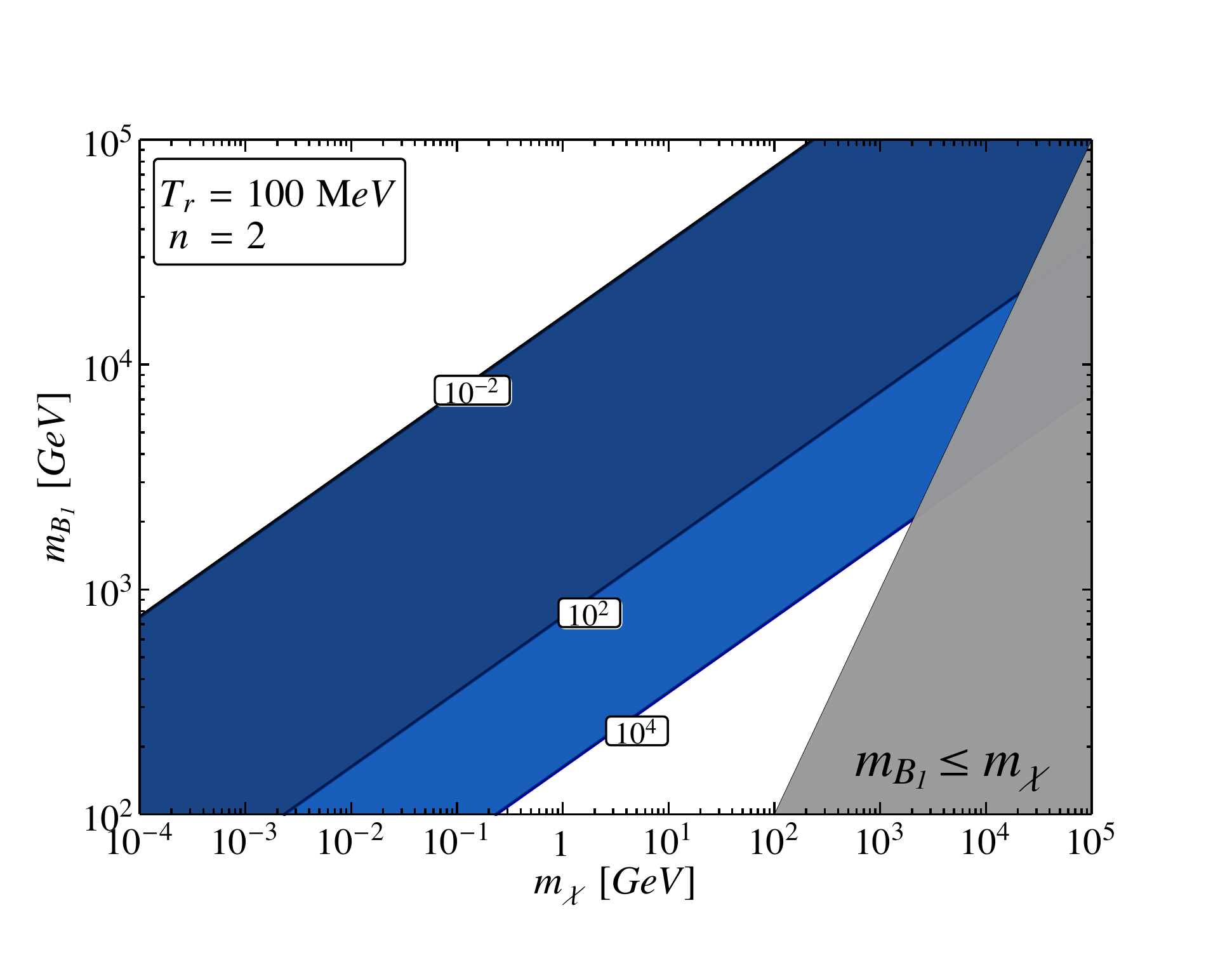}\hfill \hspace{-1.2cm}
\includegraphics[width=.282\textwidth]{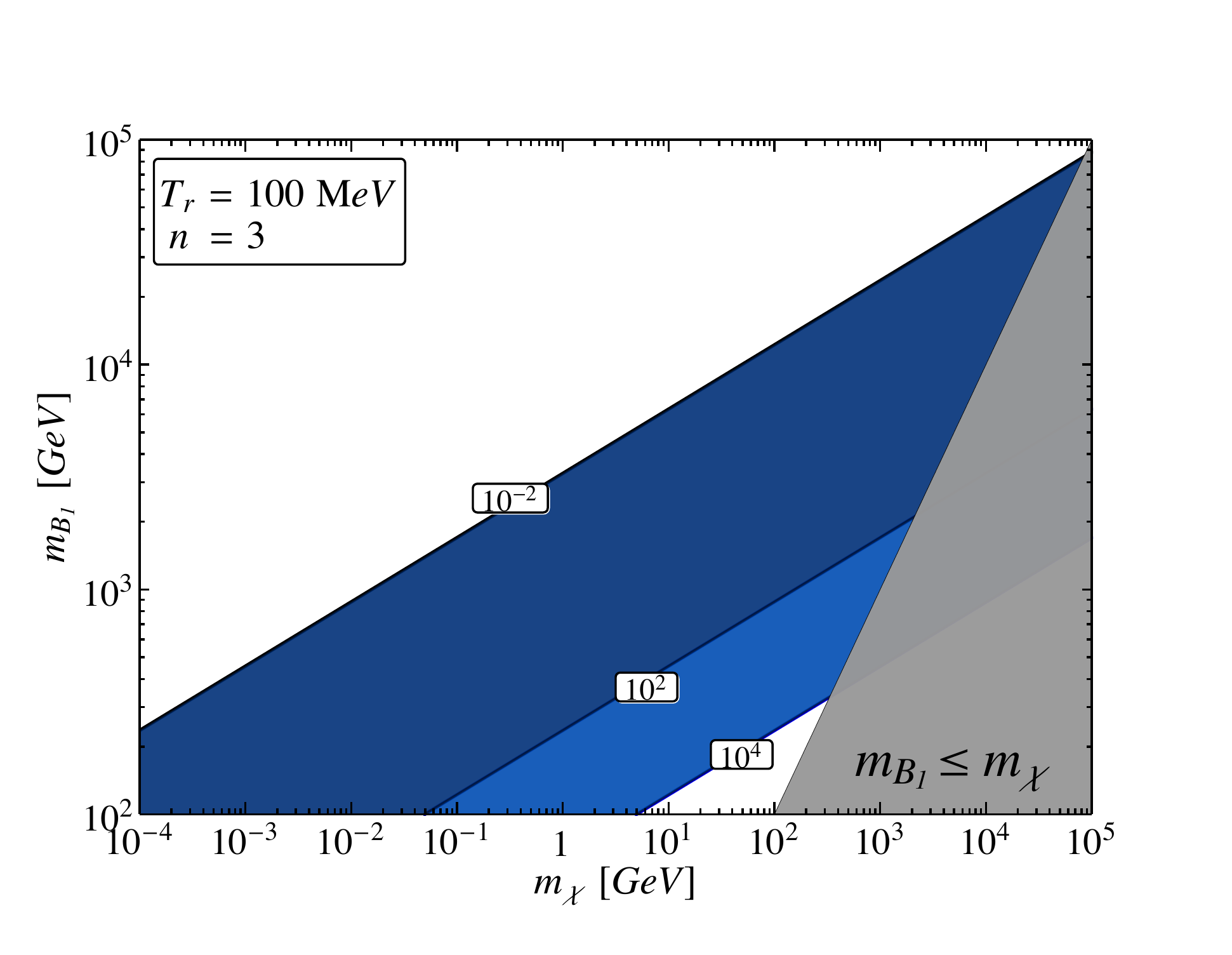}\hfill \hspace{-1.2cm}
\includegraphics[width=.282\textwidth]{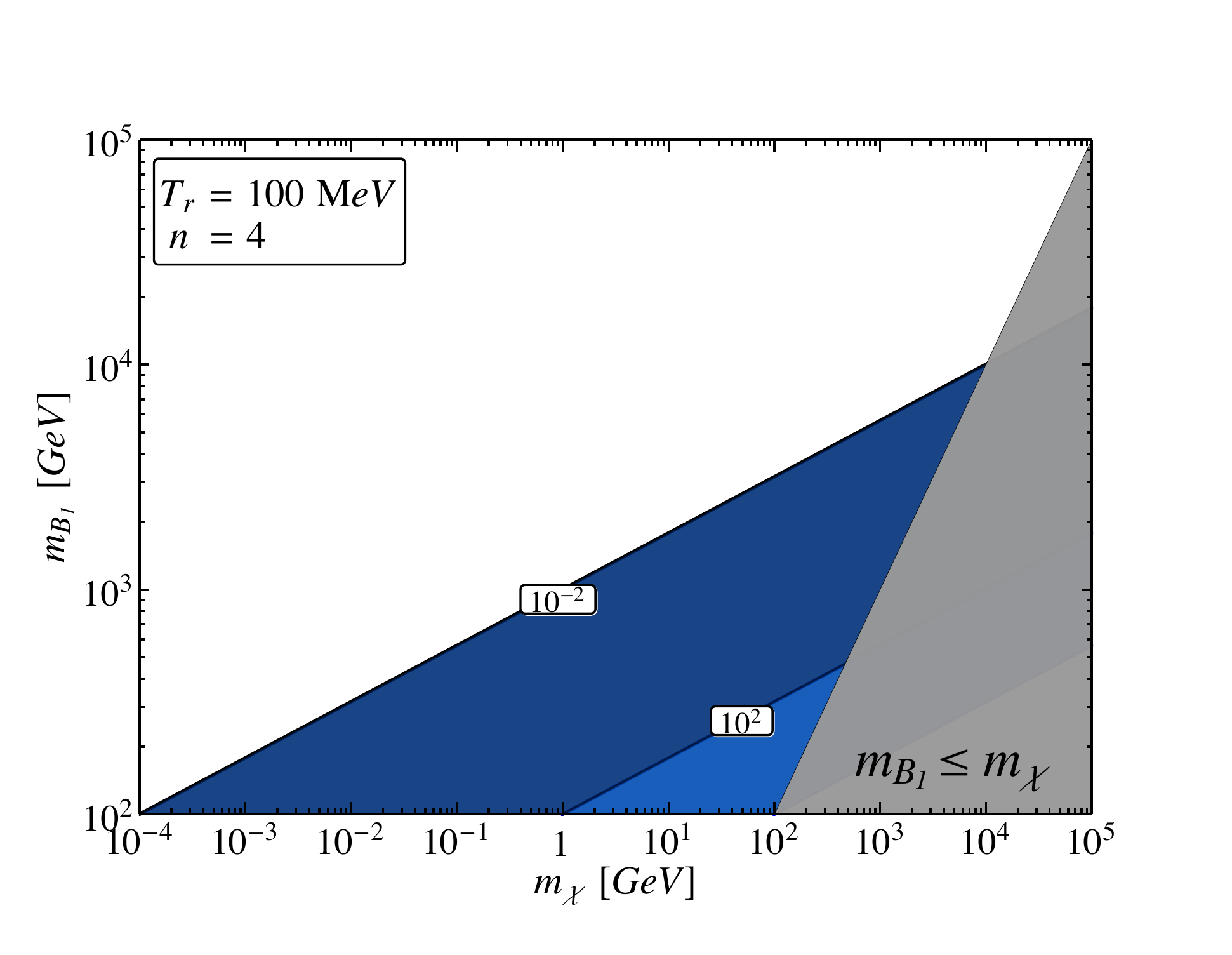}\hspace{-0.45cm} 

\caption{Contours of the $B_1$ decay length (in cm) on the $(m_{\chi}, m_{B_1})$ plane corresponding to coupling values which produce the observed DM abundance, for different values of $n$ and $T_{r}$. The blue region corresponds to $10^{2}$cm $ \leq \tau_{B_1} \leq 10^{4}$cm and the dark blue region corresponds to $10^{-2}$cm $\leq\tau_{B_1} \leq 10^{2}$cm. The first (second) row corresponds to $T_{r}=1 \gev$ ($T_{r}=100 \mev$) and the first (second, third and fourth) column corresponds to $n=1$($n=2$, $n=3$ and $n=4$). }
\label{pics:DL_mxvsmb}
\end{figure*}

\begin{figure}[htp]
\centering 
\hspace{-0.45cm}
\includegraphics[width=.283\textwidth]{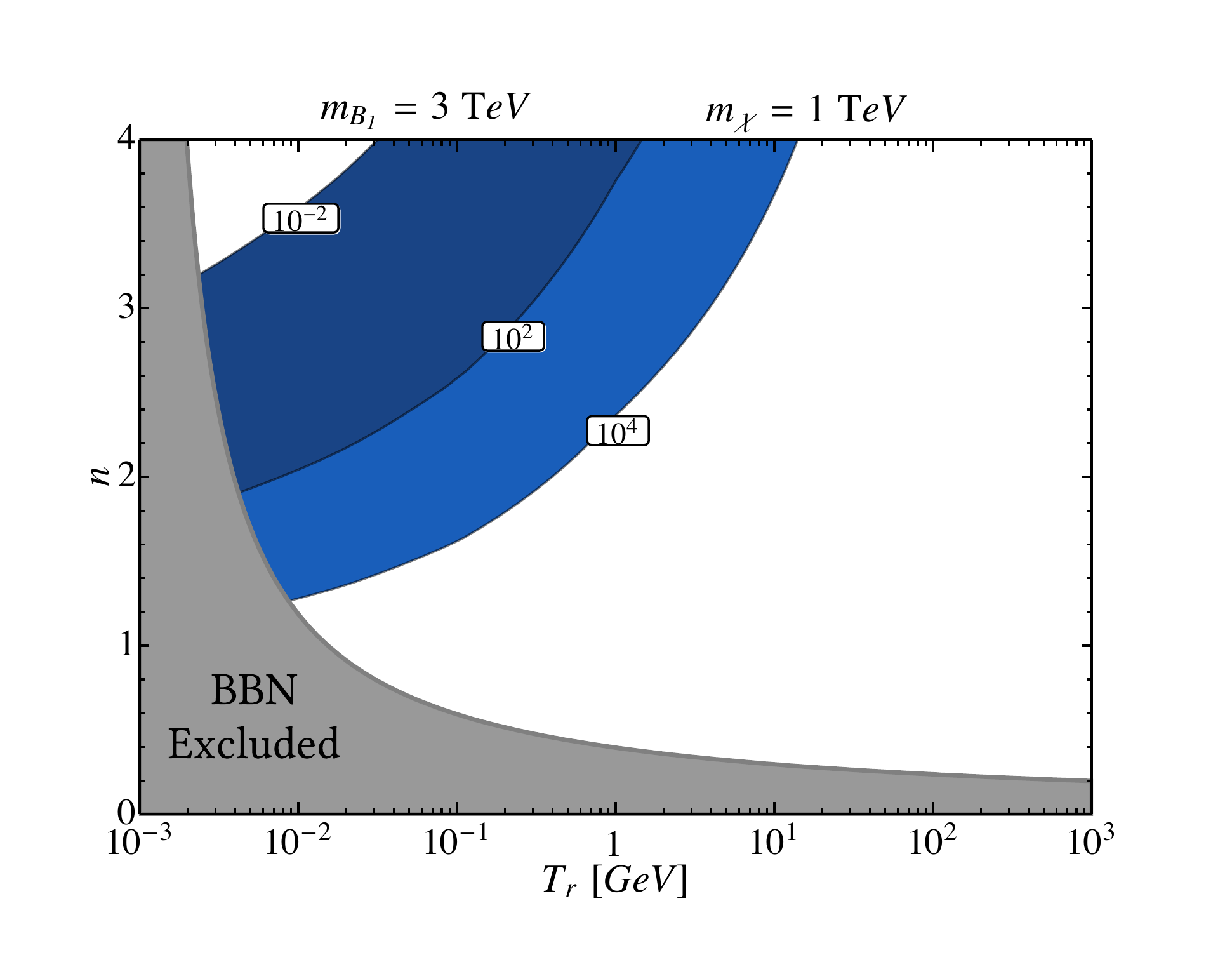}\hfill \hspace{-1.2cm}
\includegraphics[width=.283\textwidth]{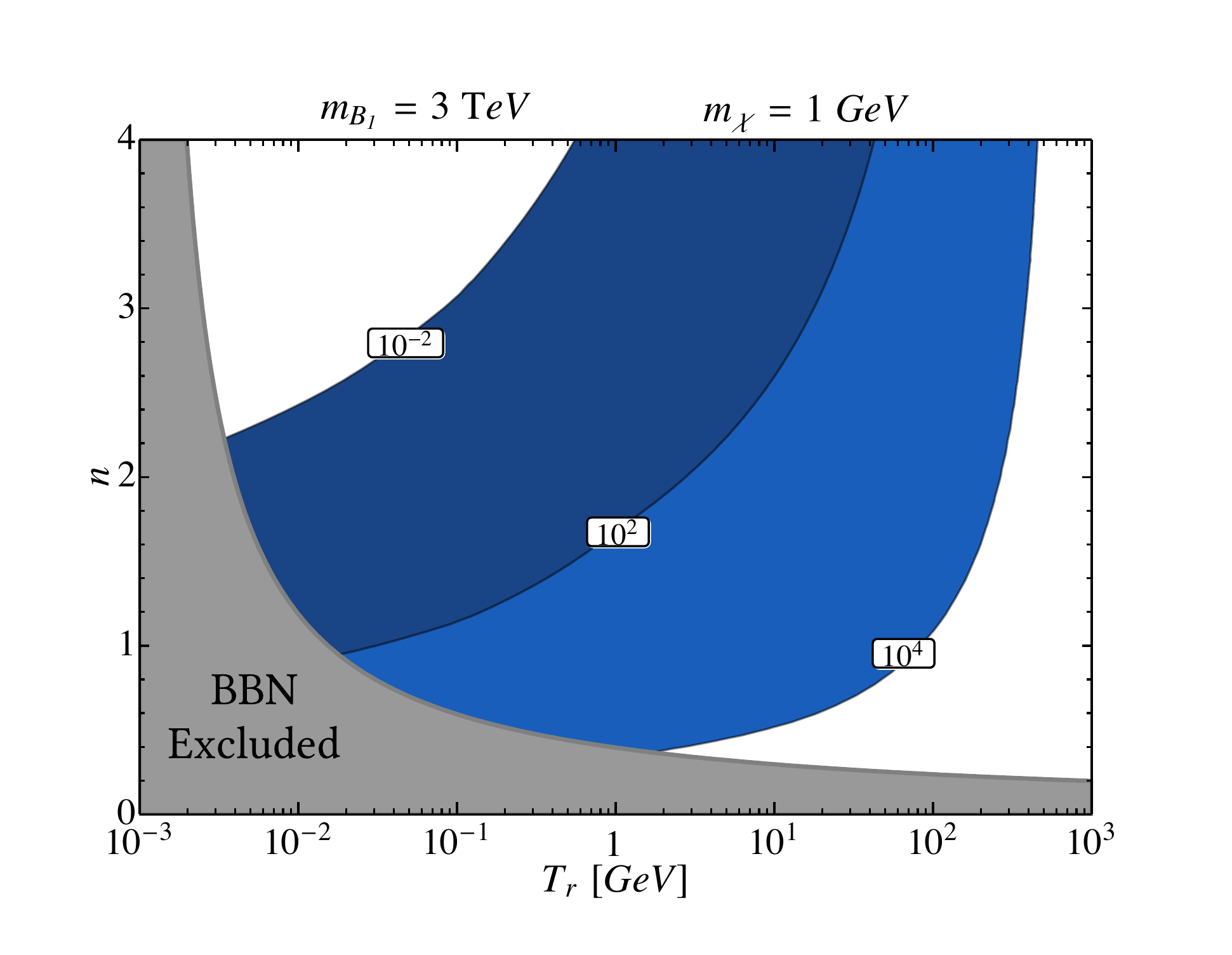}\hspace{-0.45cm}

\medskip
\vspace{-0.3cm}
\hspace{-0.45cm}
\includegraphics[width=.283\textwidth]{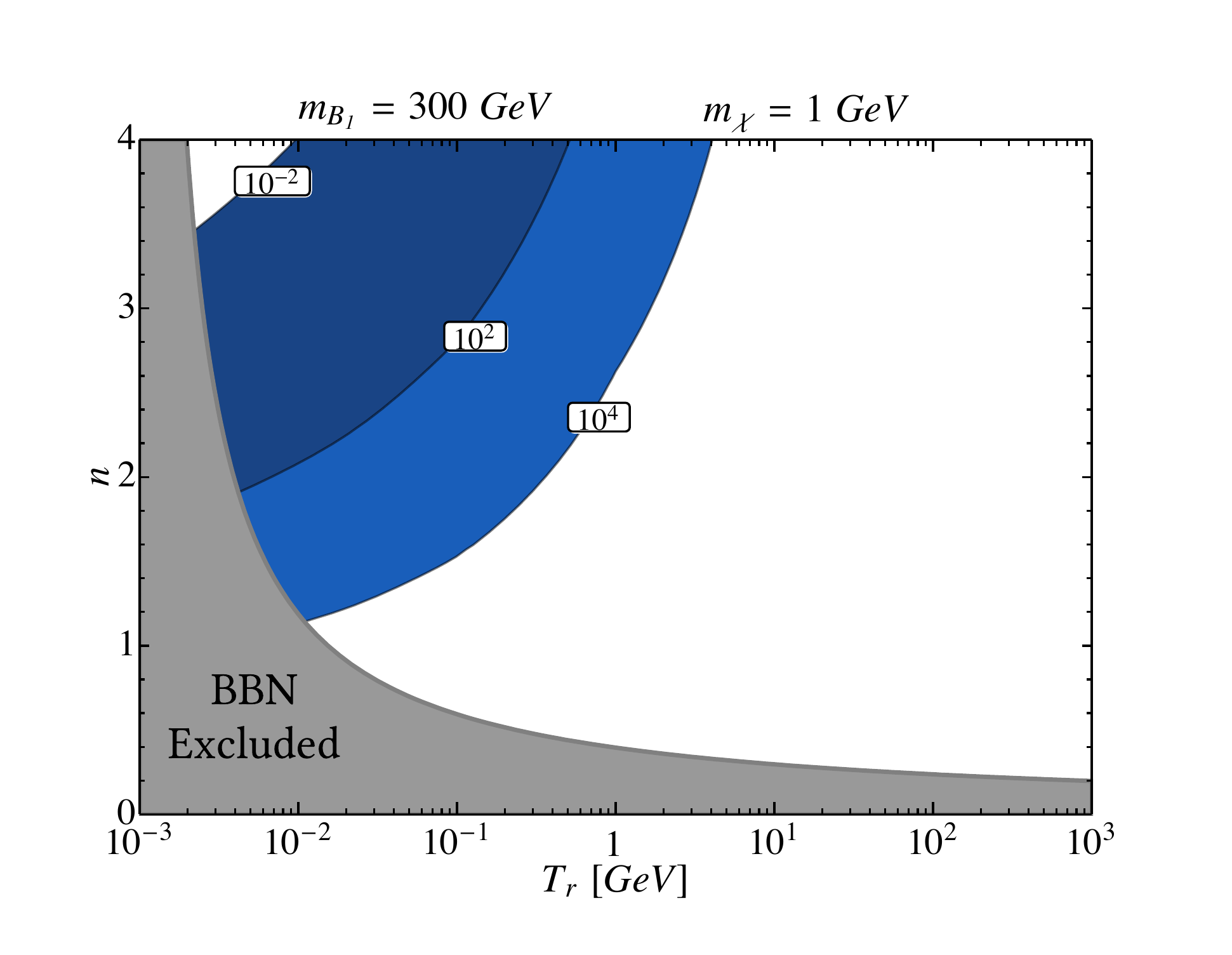}\hfill \hspace{-1.2cm}
\includegraphics[width=.283\textwidth]{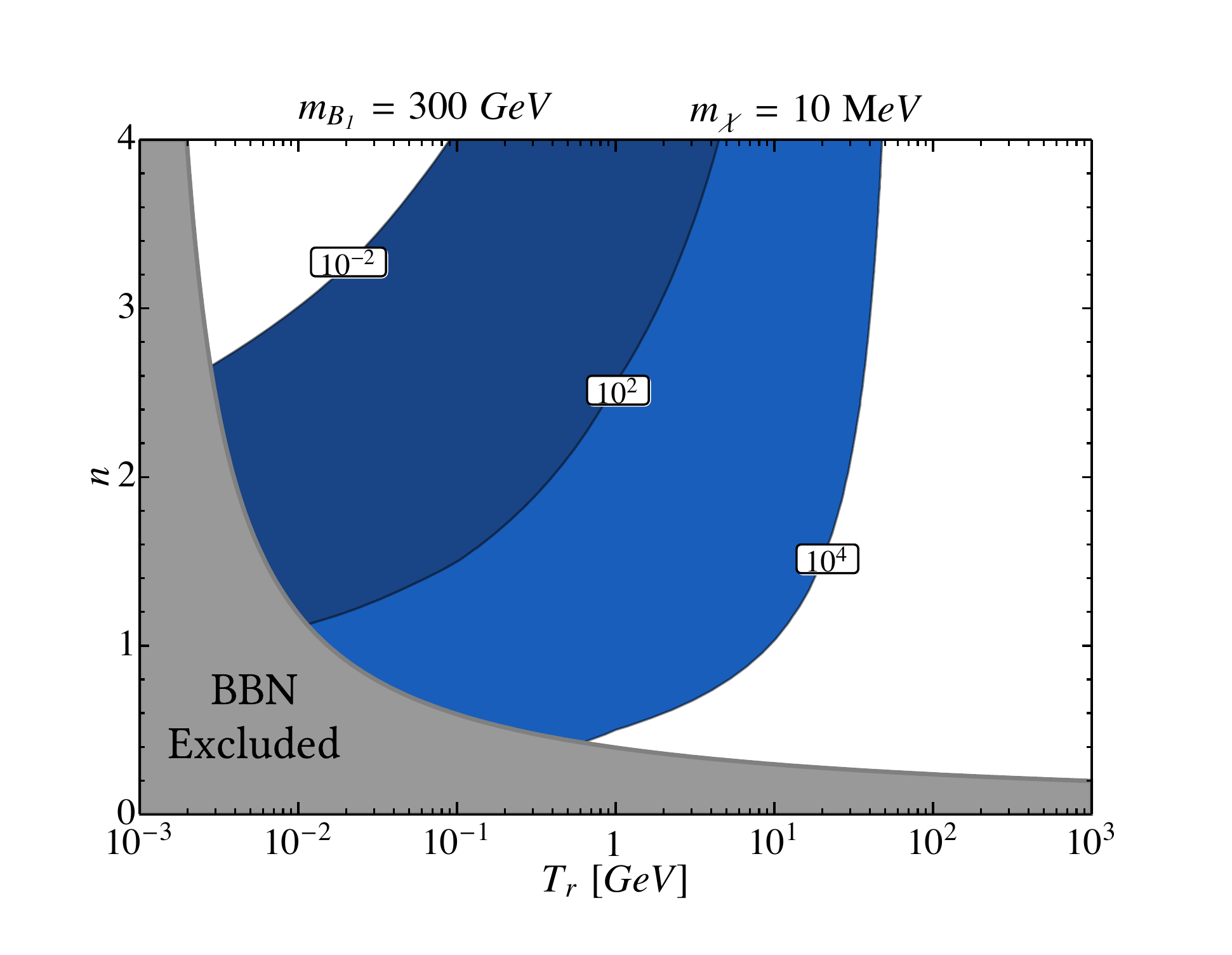}\hspace{-0.45cm} 
 
\caption{Contours of the $B_1$ decay length (in cm) on the $(T_{r},n)$ plane that reproduces the observed DM abundance for different values of $m_{B_1}$ and $m_{\chi}$. The  blue region corresponds to $10^{2}$cm $ \leq \tau_{B_1} \leq 10^{4}$cm and the dark blue region corresponds to $10^{-2}$cm $\leq\tau_{B} \leq 10^{2}$cm. We fix $m_{B_1}=3 \tev$ and $m_{B_1}=300 \gev$  for the first and second row respectively and change $m_{\chi}$ accordingly.}
\label{pics:DL_nvsT}
\end{figure}

The parameter space for displaced decays is explored in Figs. \ref{pics:DL_mxvsmb} and \ref{pics:DL_nvsT}. We start from Fig.~\ref{pics:DL_mxvsmb}, where we analyze the behavior of $\tau_{B_1}$ as we change the particle physics properties. The cosmological parameters $(T_r, n)$ are fixed for each panel, and we show the contours for $\tau_{B_1} $ on the $(m_{\chi}, m_{B_1})$ plane. The blue region corresponds to $10^{2}$ cm $ \leq \tau_{B_1} \leq 10^{4}\ \textrm{cm}$, whereas the dark blue region corresponds to $10^{-2}\ \textrm{cm}  \leq\tau_{B_1} \leq 10^{2}\ \textrm{cm}$. There are benchmarks for displaced signatures at colliders. The gray region in the bottom right corner is excluded by kinematics. We observe that isocontours follow the lines where $m_\chi \propto m_{B_1}^{2+ n/2}$, consistently with the approximate solution given in \Eq{eq:tauB1approx}.\footnote{It is important to remember that there is a power of $m_{B_1}$ in $\mathcal{F}(T_r, n)$ through $x_r$, since for each panel this time $T_r$ is the fixed quantity, see the definition in \Eq{eq:Fdecaydef}.} Moreover, we see that the decay length is reduced by a factor of $\sim (T_{r}/m_{B_1})^{n/2}$ with respect to the radiation case. For example, if we take $m_{\chi}=10\gev$ and $m_{B_1}=3 \tev$, the expected decay length for radiation case $\tau_{B_1}\sim 3 \times 10^{6} \textrm{cm}$. In our modified cosmological histories, the decay length expands into a range where its values vary from $10^{-2}\  \textrm{cm}$ to $10^{4}\  \textrm{cm}$. This range is accessible to present or future colliders.

In  Fig. \ref{pics:DL_nvsT} we study the decay length $\tau_{B_1}$ as we change the cosmological parameters, offering a complementary view of our results. The value of  $m_{B_1}$ and  $m_{\chi}$ are fixed now for each panel, and we show iso-contours for $\tau_{B_1} $ in the $(n,T_r)$ plane. The bottom left corner grey area is the region excluded by BBN.

\section{Freeze-In from Scattering}
\label{sec:scattering}

We now focus on models where DM is produced out of equilibrium via $2 \rightarrow 2$ scattering processes. As already explained in the Introduction, we divide the discussion into two classes of models, according to the number of DM particles produced for each reaction. We study the DM number density evolution for both scenarios, and we then discuss the implications for experimental searches. 

\subsection{DM Single Production}

We start our analysis from models where DM particles are produced in the early universe via bath particle scattering of the form
\be
B_1 B_2 \rightarrow B_3 \chi \ .
\label{eq:singleDM}
\ee
This is the leading production mechanism for several DM models. For example, the supersymmetric partner to the axion, the {\it axino}, in supersymmetric Peccei-Quinn theories, is a motivated DM candidate~\cite{Rajagopal:1990yx,Covi:1999ty,Covi:2001nw,Strumia:2010aa,Chun:2011zd,Bae:2011jb,Choi:2013lwa}  produced via scattering as in~\Eq{eq:singleDM}. The bath particles producing the {\it axino} depend on the specific implementation of the PQ symmetry. For KSVZ theories~\cite{Kim:1979if,Shifman:1979if}, the axino is produced via scattering of gluons and gluinos, whereas for DFSZ theories~\cite{Dine:1981rt,Zhitnitsky:1980tq} the processes can also be initiated by Higgs bosons and higgsinos. 

The general collision operator for this class of models is derived in App~\ref{app:thermalaverages}, where we find the two equivalent expressions in Eqs.~\eqref{eq:appcoll5} and \eqref{eq:appcoll6}. In our numerical analysis, we choose each time the most convenient one according to the mass spectrum of the theory (see the Appendix for details). 

We observe that the process in \Eq{eq:singleDM} is not the only channel for DM production. The two reactions obtained by taking a permutation of the bath particles are allowed by crossing symmetry, and we must account for them as well. The way crossing symmetry is implemented depends on the specific model. Here, we study benchmark models where the matrix element is left unchanged under crossing symmetry. Moreover, we assume that the matrix element for this process is independent on the kinematics. We parameterize the squared matrix elements as follows~\footnote{Notice that this happens exactly, for example, when the particles involved in the reaction are scalar fields and the interaction is of the type $\mathcal{L} = \lambda_{B \chi} B_1 B_2 B_3 \chi$.}
\be
\begin{split}
\lambda_{B \chi}^{2}  = & \, \absval{\mathcal{M}_{B_1 B_2 \rightarrow B_3 \chi}}^2 = \\ &
\absval{\mathcal{M}_{B_2 B_3 \rightarrow B_1 \chi}}^2 = \absval{\mathcal{M}_{B_1 B_3 \rightarrow B_2 \chi}}^2 \  . 
\end{split}
\ee

For models satisfying these assumptions, the collision operator takes the simple form in \Eq{eq:appcollfinal1}, which we report here in the final form
 \be
\begin{split}
\mathcal{C}_{B_i B_j \rightarrow B_k \chi} = & \, \frac{\lambda_{B \chi}^{2} \, T}{512 \pi^5} \,  
\int_{s_{\rm single}^{\rm min}}^\infty \frac{ds}{s^{3/2}} \, K_1[\sqrt{s} / T] \, \times \\ &  
\lambda^{1/2}(s, m_{B_i}, m_{B_j}) \, \lambda^{1/2}(s, m_{B_k}, m_{\chi}) \ ,
\end{split}
\label{eq:collsinglesec}
\ee
where the function $\lambda(x,y,z)$ is defined in \Eq{eq:lambdadef} and the lower integration limit is set by the kinematical threshold for the reaction
\be
s_{\rm single}^{\rm min} = {\rm max}\left\{(m_{B_i} + m_{B_j})^2 , (m_{B_k} + m_\chi)^2 \right\} \ .
\ee

We analyze the number density evolution for the class of models introduced above. We fix the masses to be $(m_{B_{3}}, m_{\chi}) = (1000, 10) \gev$, whereas bath particles $B_1$ and $B_2$ have negligible masses. As an example, this is the case where $\chi$ is the axino, $B_3$ is the gluino and $B_{1,2}$ are gluons. For this choice of the parameters, the observed DM  abundance is reproduced for the standard cosmology if we choose $\lambda^{rad}_{B \chi} = 1.5 \times 10^{10}$.

\begin{figure}[htp]
\center\includegraphics[width=0.483 \textwidth]{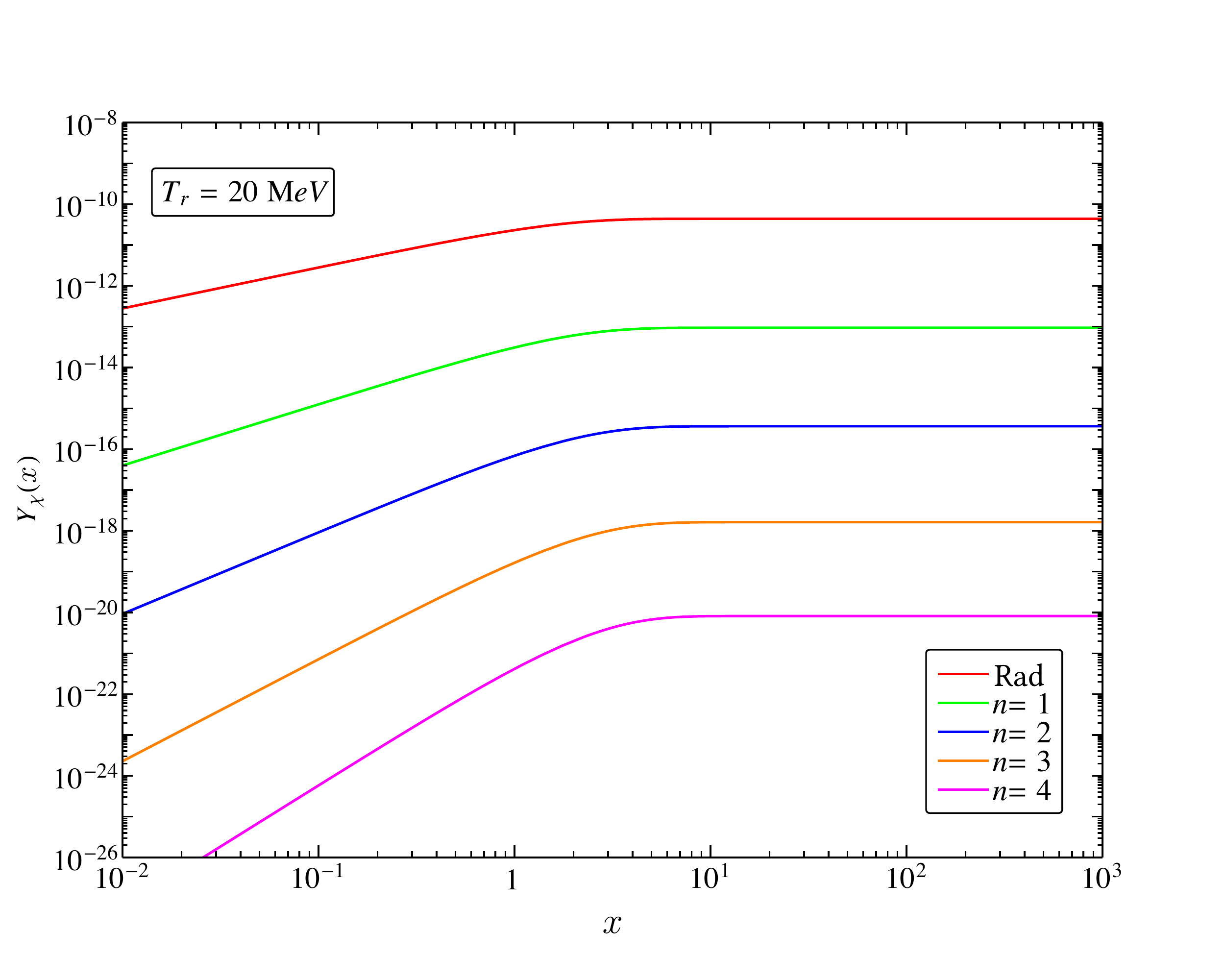}
\caption{Numerical solutions for the comoving number density $Y_{\chi}$ as a function of $x = m_{B_3} / T$. We choose $m_{B_{3}} = 1\ \tev$, $m_{\chi}=10\ \gev$ and $\lambda_{B \chi}= 1.5 \times 10^{-10}$. We consider different values for $n$, whereas we always set $T_{r} = 20 \mev$.}
\label{fig:FI_Scattering_Mb1000_Mx10}
\end{figure}

Numerical results for the number density evolution are shown in Fig.~\ref{fig:FI_Scattering_Mb1000_Mx10}, where we set $T_{r} = 20 \mev$ and we consider a few different values of $n$ as indicated in the caption. We plot the solution as a function of the ``time variable'' $x = m_{B_3} / T$. The behavior is similar to the one already seen for freeze-in via decays: the asymptotic comoving density decreases as we increase the value of the index $n$. The net effect is that DM is underproduced, which in turn requires larger cross sections to reproduce the observed DM abundance. The asymptotic number density is reached for $x \simeq 4$, or equivalently for temperatures $T \simeq m_{B_3} / 4$. This is not surprising, since $B_3$ is the heaviest particle involved in the reaction. In order to produce a DM particle, we either need a $B_3$ particle in the initial state or enough kinetic energy to produce $B_3$ in the final state. At the temperature drops below $m_{B_3}$, these processes become exponentially rare. 

As in the decays of freeze-in via decays, the behavior of the numerical solutions can be reproduced analytically. In order to do so, we only keep  the finite mass of $B_3$. This is well justified for the spectrum under consideration: $m_{B_3} = 100 m_\chi$, whereas $B_1$ and $B_2$ are massless. Once we make this approximation, the collision operator in \Eq{eq:collsinglesec} simply reads
\be
\begin{split}
\mathcal{C}_{B_i B_j \rightarrow B_k \chi} = & \, \frac{\lambda_{B \chi}^{2} \, T}{512 \pi^5} \,  
\int_{m_{B_3}^2}^\infty ds \, \frac{s - m_{B_3}^2}{s^{1/2}} K_1[\sqrt{s} / T]  = \\ &
 \frac{\lambda_{B \chi}^{2} \, m_{B_3}^4}{128 \pi^5} \, \frac{K_1[x]}{x^3}  \ ,
\end{split}
\ee
where we remind the Reader that $x = m_{B_3} / T$. This result is valid for any permutation of the bath particles, thus the total collision operator is obtained by multiplying the above result by a factor of three. 

The freeze-in yield is obtained from the general result in \Eq{eq:BEforFI4}. Upon neglecting as usual the temperature variation of $g_*$, we find the approximate solution
\be
Y_\chi(x) \simeq \frac{\lambda^{2}_{B \chi}}{\overline{g}^{3/2}_*} \, \frac{405 \sqrt{10}}{256 \pi^8} \, \frac{M_{\rm Pl}}{m_{B_3} x_r^{n/2}}  \int_0^x dx^\prime K_1[x^\prime] \, x^{\prime \,(1 + n/2)}   \ ,
\label{eq:YFIscatBxapprox}
\ee
where, in this case, $x_{r} \approx m_{B_{3}}/T_{r}$. Considering early times, $x \ll 1$, we can Taylor-expand the Bessel function and calculate the slope of the lines in Fig.~\ref{fig:FI_Scattering_Mb1000_Mx10}
\be
Y_\chi(x) \propto x^{(1 + n/2)} \qquad \qquad (x \ll 1) \ .
\label{eq:slopeDMsingle}
\ee
This scaling is different from the result we found for decays in \Eq{eq:FIscaling}. Consistently, the slopes of the curves in Fig.~\ref{fig:FI_Decay} and Fig.~\ref{fig:FI_Scattering_Mb1000_Mx10} are different.

The asymptotic value for the yield can be computed by evaluating the approximate solution in \Eq{eq:YFIscatBxapprox} for $x \rightarrow \infty$. As done before, it is convenient to normalize our solution with respect to the result in a radiation-dominated ``standard'' early universe
\be
\left. Y_\chi^\infty\right|_{\rm rad} = \frac{\lambda^{2}_{B_3 \chi}}{\overline{g}^{3/2}_*} \, \frac{405 \sqrt{10}}{512 \pi^7} \,  \frac{ \, M_{\rm Pl}}{m_{B_3}}  \ .
\ee
We express the asymptotic value as it follows
\be
Y_\chi^\infty  = \left. Y_\chi^\infty\right|_{\rm rad} \times  \mathcal{F}_{{\rm scatt}}^{B \chi} (T_r, n) \ , 
\label{eq:YchiscatBxanalytical}
\ee
where we define the function
\be
\mathcal{F}_{{\rm scatt}}^{B \chi} \equiv \frac{2}{\pi}\left(\frac{2}{x_r}\right)^{n/2} \varGamma\left[\frac{2 + n}{4}\right]  \varGamma\left[\frac{6 + n}{4}\right]  \ .
\label{eq:FscattBxdef}
\ee

\begin{figure}[t]
\center\includegraphics[width=0.483 \textwidth]{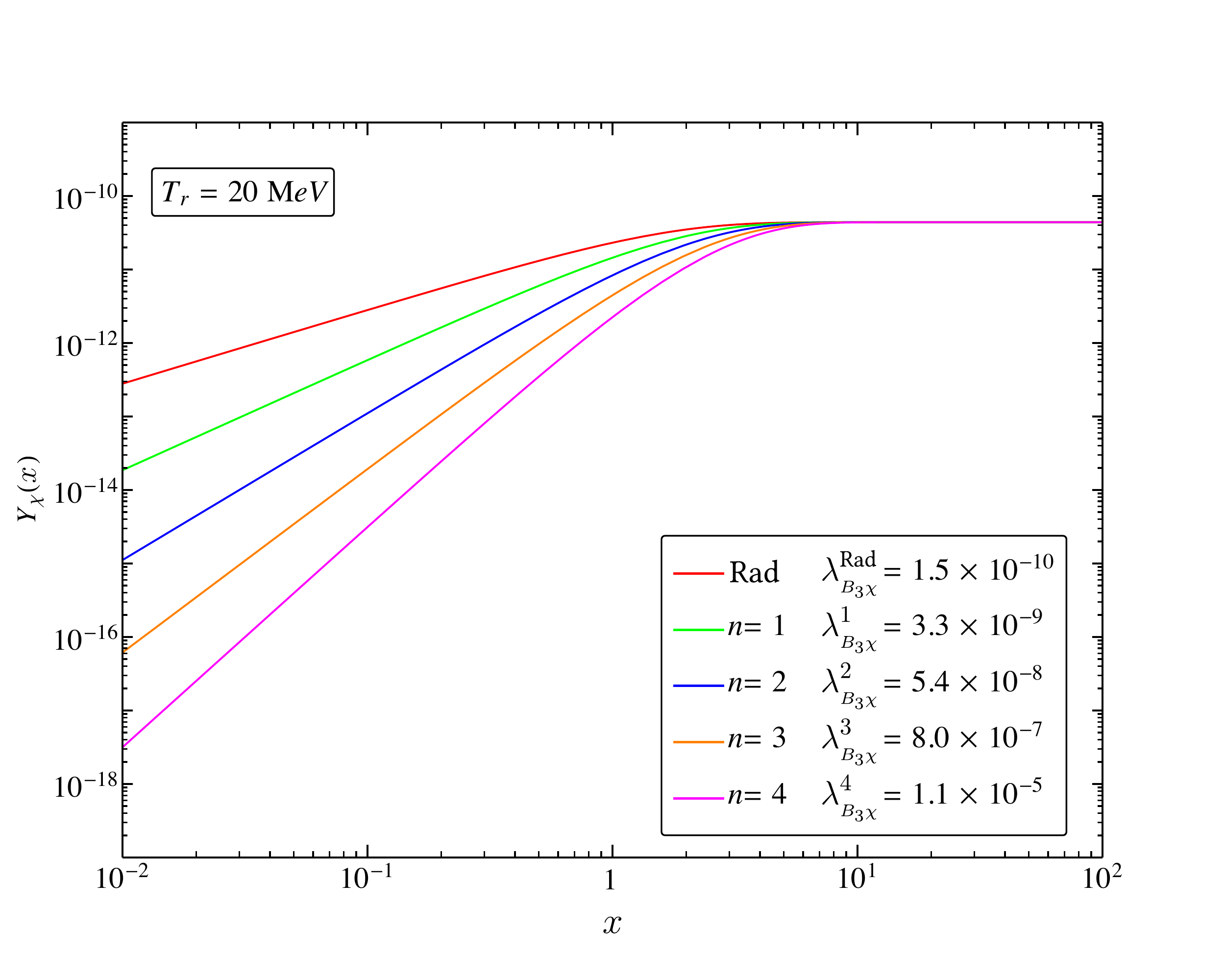}
\caption{Numerical solution for the comoving number density $Y_\chi$ with $m_{B_{3}}=1\tev$ and $m_{\chi}=10\gev$. Now $\lambda_{B_3 \chi}$ is changed in order to reproduce the observed abundance ($\lambda^{1}_{B_3 \chi}=3.3 \times 10^{-9}$, $\lambda^{2}_{B_3 \chi}=5.4 \times 10^{-8}$, $\lambda^{3}_{B_3 \chi}=8.0 \times 10^{-7}$, $\lambda^{4}_{B_3 \chi}=1.1 \times 10^{-5}$).  We set $T_{r}=20 \mev$ for all $n$.}
\label{fig:FI_ScattSingle_samerelic_Mb1000_Mx10}
\end{figure}

In Fig.~\ref{fig:FI_ScattSingle_samerelic_Mb1000_Mx10}, we use the same mass and $T_r$ values, but we choose this time $\lambda_{B_{3} \chi}$ to reproduce the observed DM abundance for each $n$. The enhancement in the matrix element can be as large as $\sim 10^{5}$. Such enhancements for the couplings translate into a quadratically larger effect in the cross sections for potential DM detection processes, which can be enhanced by a factor of $10^{10}$.

\begin{figure}[tp]
\center\includegraphics[width=0.483 \textwidth]{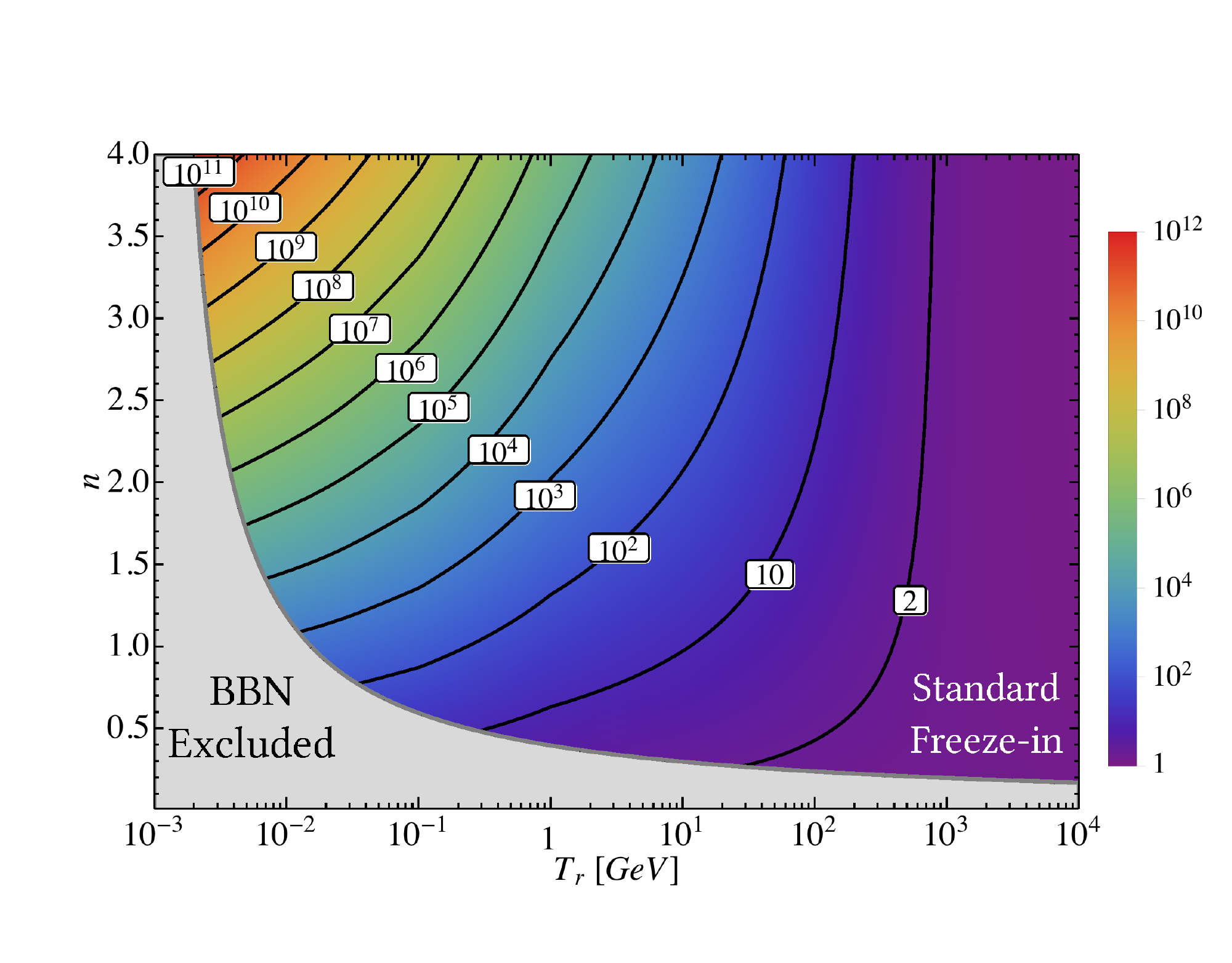}
\caption{Reduction in the relic density in the case of Freeze-in by scattering $B_1 B_2 \rightarrow B_3 \chi$ for $m_{B_{3}}=1\tev$ and $m_{\chi}=10\gev$ compare with the observed DM density in the standard case (radiation)}
\label{fig:FI_Scattering_E_Mb1000_Mx10}
\end{figure}

For freeze-in via scattering $B_1 B_2 \rightarrow B_3 \chi$, the DM relic abundance is always suppressed in the $(T_r, n)$ plane. We quantify this suppression in Fig.~\ref{fig:FI_Scattering_E_Mb1000_Mx10}, where we keep the same mass values for $B_3$ and $\chi$. More specifically, we show iso-contours of the function
\beq
r_{B\chi}(T_r, n) \equiv \dfrac{\left.\Omega_\chi h^2\right|_{\rm rad}}{\Omega_\chi h^2}  \ . 
\eeq

The suppression factor can be analytically understood by using the equations derived above
\beq
r_{B\chi}(T_r, n) \simeq \mathcal{F}_{{\rm scatt}}^{B \chi} (T_r, n)^{-1} \simeq \left(\frac{x_r}{2}\right)^{n/2} \,.
\eeq
The associated enhancement in the required matrix element $\lambda_{B_3 \chi}$ reads 
\beq
\lambda_{B \chi}(T_r, n) = r_{B\chi}(T_r, n)^{1/2} \, \lambda_{B \chi}^{\rm rad} \ ,
\eeq
 indicating that increasing $n$ and/or decreasing $T_{r}$ leads to larger values for the necessary coupling constant to reproduce the observed DM density of the universe.

\subsection{DM Pair Production}

We consider in this section the third and final freeze-in case: DM pair production
\be
B_1 B_2 \rightarrow \chi \chi \ .
\label{eq:pairDM}
\ee
This process is the leading production mechanism for all models where the DM particle belongs to a dark sector very weakly coupled to the visible sector. Notable examples include Higgs portal models with small mixing angle and dark photon models with small kinetic mixing.

General results for this case are also given in App~\ref{app:thermalaverages}, where the two equivalent forms are in Eqs.~\eqref{eq:appcollpair1} and \eqref{eq:appcollpair2}. We focus also for this case on models where the matrix element is a constant
\be
\lambda_{\chi \chi}^{2} = \absval{\mathcal{M}_{B_1 B_2 \rightarrow \chi \chi}}^2 \ .
\ee
The collision operator then takes the simple form in \Eq{eq:appcollfinal2}. We write it here in the final form
\be
\begin{split}
\mathcal{C}_{B_1 B_2 \rightarrow \chi \chi} = & \, \frac{\lambda_{\chi \chi}^{2} \, T}{512 \pi^5} \,  
\int_{s_{\rm pair}^{\rm min}}^\infty \frac{ds}{s^{3/2}} \, K_1[\sqrt{s} / T] \, \times \\ &  
\lambda^{1/2}(s, m_{B_1}, m_{B_2}) \, \lambda^{1/2}(s, m_\chi, m_{\chi}) \ ,
\end{split}
\label{eq:collpairsec}
\ee
with $\lambda(x,y,z)$ defined in \Eq{eq:lambdadef} and the lower integration limit set by kinematics
\be
s_{\rm pair}^{\rm min} = {\rm max}\left\{(m_{B_1} + m_{B_2})^2 , (2 m_\chi)^2 \right\} \ .
\ee

\begin{figure}[tp]
\center\includegraphics[width=0.483 \textwidth]{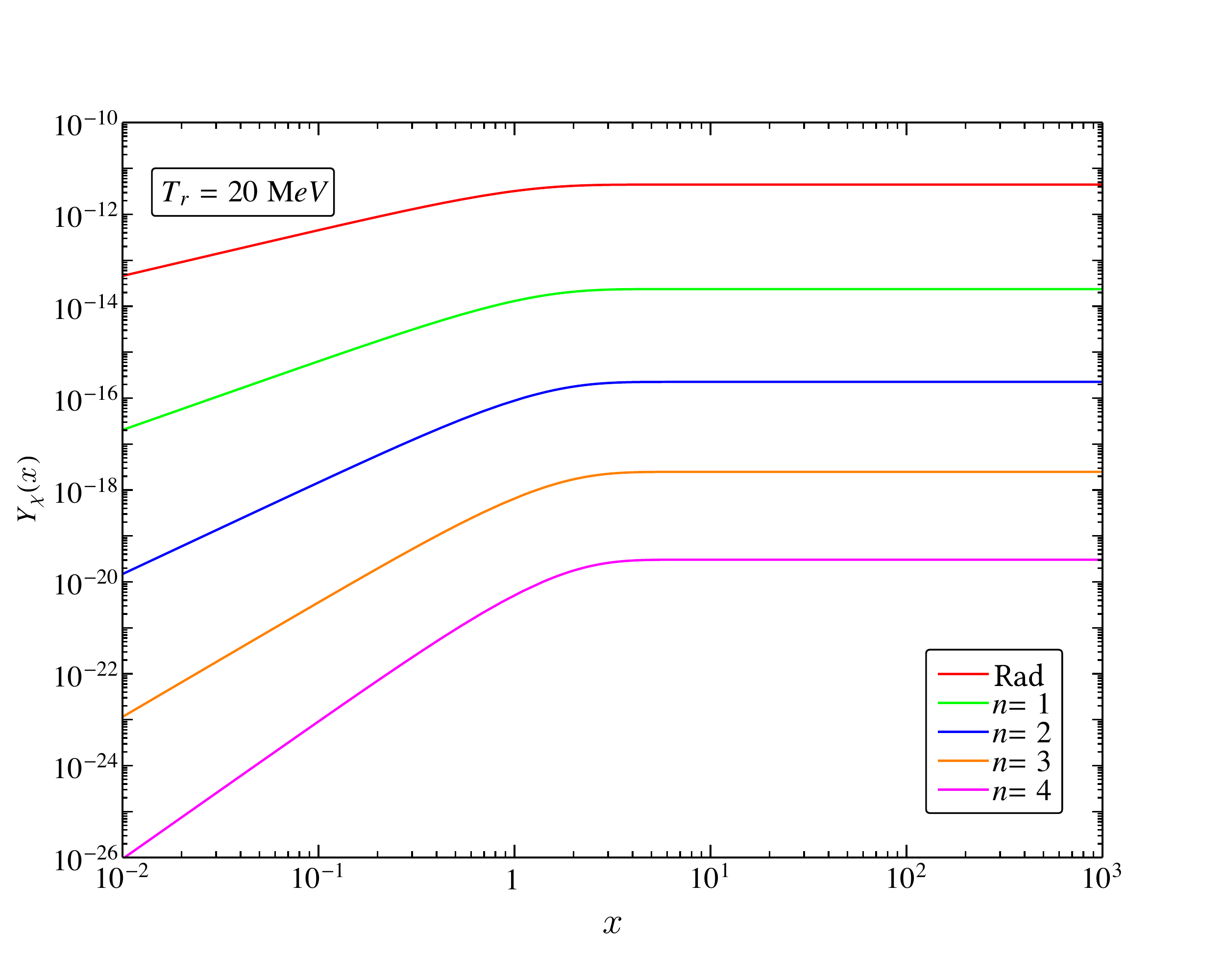}
\caption{Numerical solution for  the comoving number density $Y_{\chi}$ with  $m_{\chi}=100\gev$ and $\lambda_{\chi\chi}= 3.41 \times 10^{-11}$ in the case of Freeze-in by scattering $B_1 B_2 \rightarrow \chi \chi$. We set $T_{r} = 20 \mev$ for all $n$.}
\label{fig:FI_Scattering_xx}
\end{figure}

\begin{figure}[tp]
\center\includegraphics[width=0.483 \textwidth]{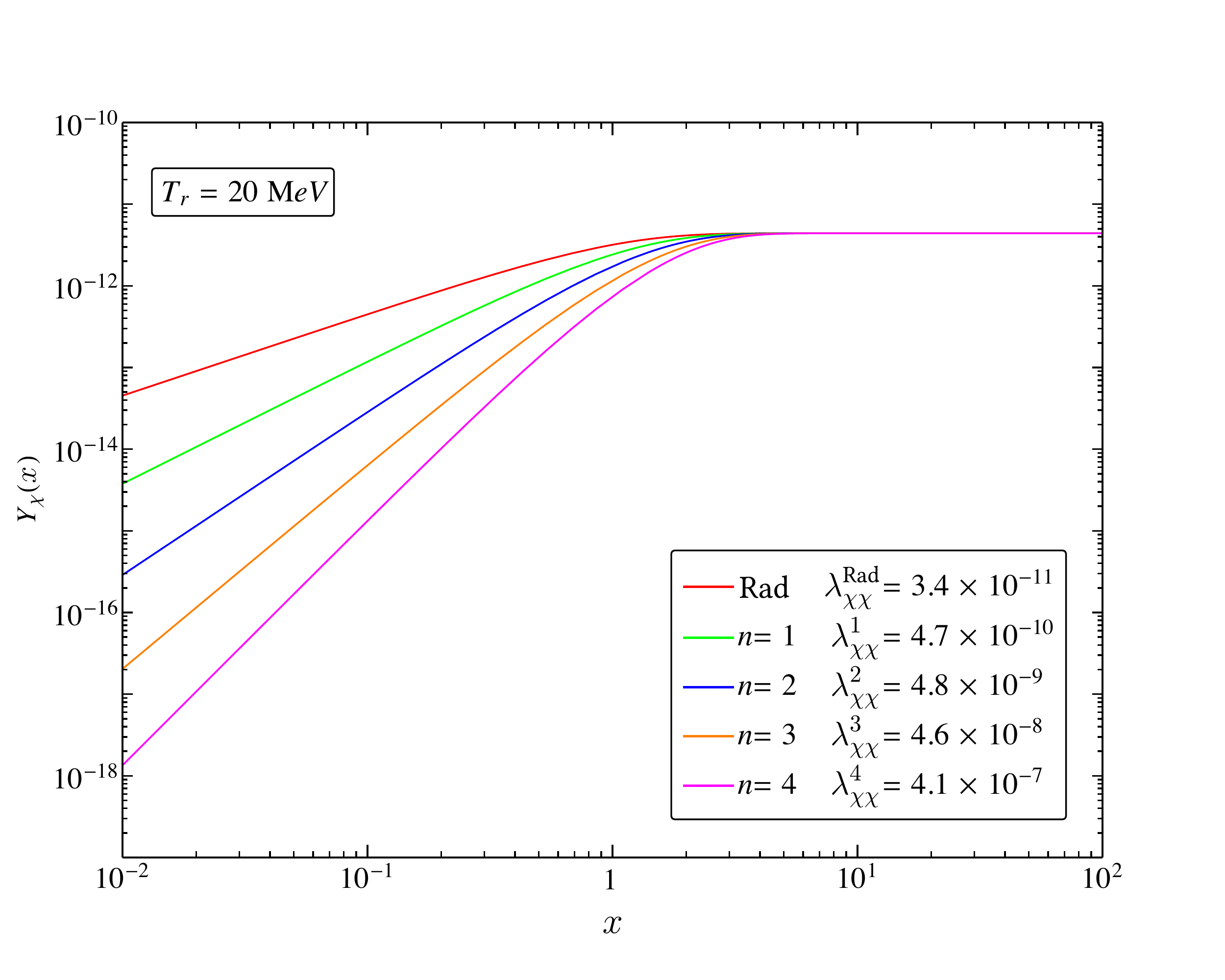}
\caption{Numerical solution for the comoving number density $Y_\chi$ with $m_{\chi}=100\gev$. Now $\lambda_{\chi \chi}$ is changed in order to reproduce the observed abundance ($\lambda^{1}_{\chi \chi}=4.7 \times 10^{-10}$, $\lambda^{2}_{\chi \chi}=4.8 \times 10^{-9}$, $\lambda^{3}_{\chi \chi}=4.6 \times 10^{-8}$, $\lambda^{4}_{\chi \chi}= 4.1 \times 10^{-7}$).  We set $T_{r}=20 \mev$ for all $n$.}
\label{fig:FI_Scattering_xx_samerelic}
\end{figure}

Numerical results for the comoving yield are presented in Fig.~\ref{fig:FI_Scattering_xx}. We choose $m_\chi = 100 \, {\rm GeV}$, we neglect the bath particle masses and we set $\lambda_{\chi\chi} = \lambda^{\rm rad}_{\chi\chi} = 3.41 \times 10^{-11}$. This is the value that reproduces the correct abundance for a standard cosmological history. We plot the comoving number density as a function of the ``time variable'' $x= m_\chi / T$. The freeze-in abundance is largely suppressed compared to the standard case also for DM pair production, forcing markedly larger couplings to explain the observed abundance. Quantitatively, this is illustrated in Fig.~\ref{fig:FI_Scattering_xx_samerelic}, where we keep $m_\chi$ and  $T_r$ fixed and we set the couplings $\lambda^n_{\chi\chi}$ needed for a modified cosmology featuring a given $n>0$. The figure shows how the needed couplings are larger than in the standard case by up to more than four orders of magnitude, for large $n\sim4$.

The analytical estimates are analogous to the previous case, and we therefore only quote the final results here. First, the collision operator neglecting the bath particles mass reads
\be
\mathcal{C}_{B_1 B_2 \rightarrow \chi \chi} = \frac{\lambda_{\chi \chi}^{2} \, m_\chi^4}{128 \pi^5} \,  \frac{K_1[x]^2}{x^2} \ .
\ee
The comoving density as a function of the temperature is obtained by computing the integral
\be
\begin{split}
Y_\chi(x) \simeq & \, \frac{\lambda^{2}_{\chi \chi}}{\overline{g}^{3/2}_*} \, \frac{135 \sqrt{10}}{256 \pi^8} \, \frac{M_{\rm Pl}}{m_{\chi} x_r^{n/2}}  \times \\ & \int_0^x dx^\prime K_1[x^\prime]^2 \, x^{\prime \,(2 + n/2)}  \ .
\end{split}
\label{eq:YFIscatxxapprox}
\ee
The slope of the different curves is the same as the one found for single DM production (see \Eq{eq:slopeDMsingle}).

We normalize again the asymptotic value with respect to the radiation case
\be
Y_\chi^\infty  = \left. Y_\chi^\infty\right|_{\rm rad} \times  \mathcal{F}_{{\rm scatt}}^{\chi \chi}\ , 
\label{eq:Ychiscatxxanalytical}
\ee
which in this case it reads
\be
\left. Y_\chi^\infty\right|_{\rm rad} = \frac{\lambda^{2}_{\chi \chi}}{\overline{g}^{3/2}_*} \, \frac{405 \sqrt{10}}{8192 \pi^6} \,  \frac{M_{\rm Pl}}{m_{\chi}}  \ .
\ee
The suppression we find in this case reads
\beq
\mathcal{F}_{{\rm scatt}}^{\chi \chi} \equiv \frac{8}{3\sqrt{\pi}} \dfrac{\varGamma\left[\frac{2 + n}{4}\right]  \varGamma\left[\frac{6 + n}{4}\right] \varGamma\left[\frac{10 + n}{4}\right]}{ x_{r}^{n/2} \varGamma\left[\frac{8 + n}{4}\right] }   \ .
\label{eq:Fscatxxdef}
\eeq

\begin{figure}[t]
\center\includegraphics[width=0.483 \textwidth]{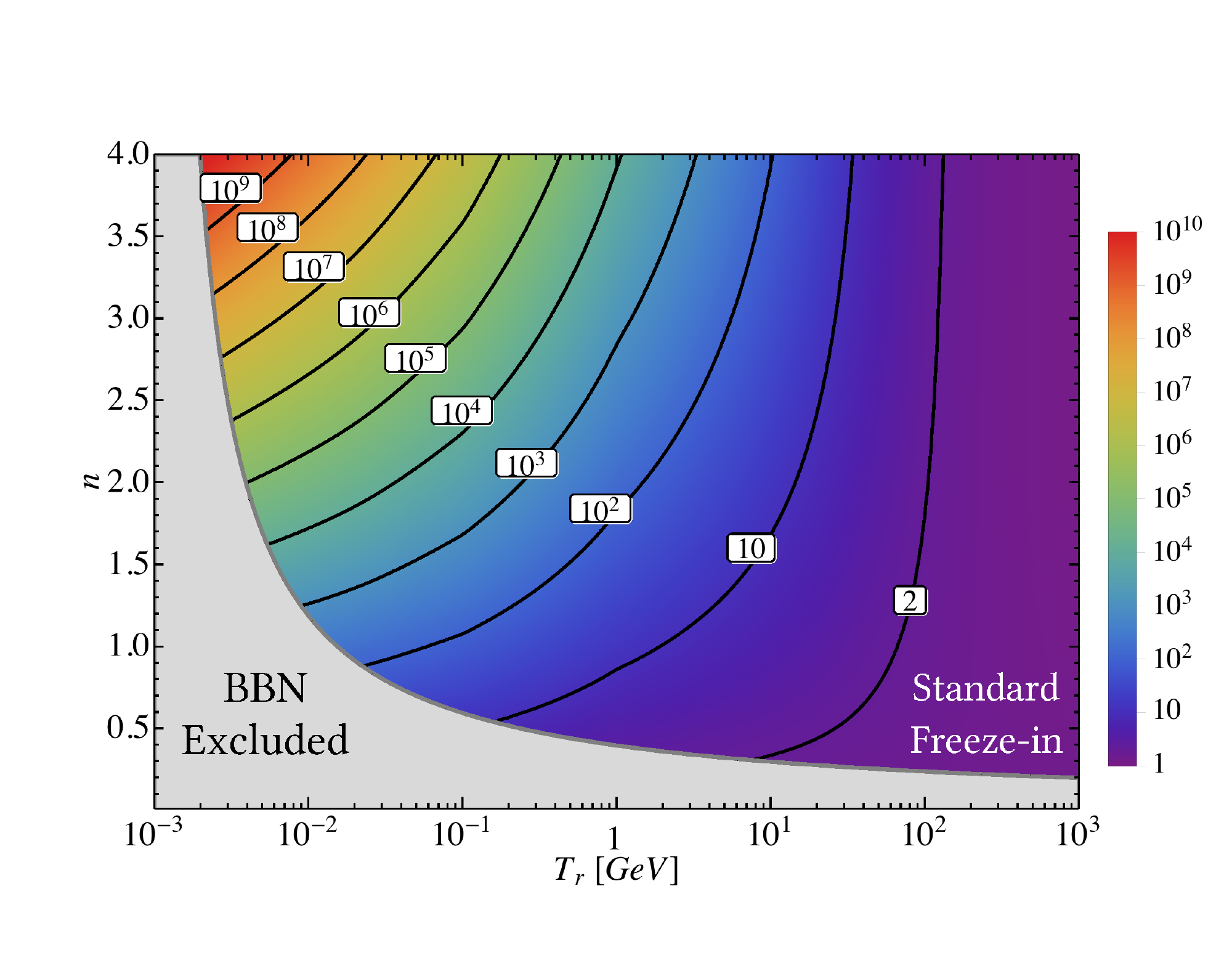}
\caption{Relic density suppression in the case of freeze-in by scattering $B_1 B_2 \rightarrow \chi \chi$, compared with the observed DM density in the standard case (radiation)}
\label{fig:FI_Scattering_xx_E}
\end{figure}

We quantify the DM relic abundance suppression $r_{\chi \,\chi} $ in the $(T_{r}, n)$ plane in Fig.~\ref{fig:FI_Scattering_xx_E} for the same values of the DM particle mass. Also in this case, suppression factors can be as large as ten orders of magnitude.

\subsection{Implications for Dark Matter Detection}

Our general finding is that when the universe is dominated, at the time of DM production through freeze-in, by a species that produces a larger Hubble rate a given temperature than in the radiation-dominated case (what we dub a ``fast-expanding'' universe), the couplings needed to produce the observed DM abundance are {\it larger} than in a radiation-dominated, standard scenario. As a result, quite generically, DM detection prospects improve.

Besides the general conclusion above, it is however hard to solidly quantify how DM detection prospects are affected in a general, model-independent way for freeze-in via scattering in modified, fast-expanding cosmologies. A first difficulty stems from the impossibility of performing a cross-symmetry prediction for, e.g., the cross section for the $B_1B_2\to \chi\chi$ process versus the cross-symmetric $\chi B_1\to \chi B_2$ process (and similarly for the single-production scattering case).

With that caveat in mind, however, for simple instances where for example the matrix element squared is a constant, as we considered above, we can attempt to draw a few general statements. Let us consider first the DM single-production case, $B_1B_2\to B_3\chi$. In this case, let us assume that for instance $B_3$ is some visible-sector species which is abundant in the late universe, for instance an electron or a photon. As long as the inverse reaction to the process leading to freeze-in $\chi$ production is kinematically allowed for non-relativistic processes, i.e., approximately,
\be
m_\chi + m_{B_3} > m_{B_1} + m_{B_2} \equiv m_{12} \ ,
\ee
and as long as $\chi$'s stability is not jeopardized by decays to $B_1+B_2+B_3$, i.e. 
\be
m_\chi - m_{B_3}  < m_{12} \ ,
\ee
{\it then}, the reaction 
\be
B_3 \chi \rightarrow B_1 B_2 
\ee
would be generically allowed, leading to potential completely novel indirect detection signals, involving a {\it single} DM particle in the initial state. Rates for this type of reaction are much larger in the fast-expanding universes we entertain here. If $B_3$ is a particle species abundant in direct detection targets, the reaction $B_3 \chi \rightarrow B_1 B_2$ would also possibly produce striking signals at direct detection experiments.

In the case of DM pair production, $B_1B_2\to \chi\chi$, and again assuming a simple form for the matrix element squared, modified cosmologies would give a strikingly large enhancement to late-time DM pair annihilation rates, $\chi\chi\to B_1B_2$. The relevant pair-annihilation cross sections, however, would presumably be quite small, unless $m_\chi\ll \sim$ GeV, since one would naively estimate, given what we find above,
\be
\sigma_{\chi\chi\to B_1B_2}\sim \frac{\lambda^2_{\chi\chi}}{m_\chi^2}\gtrsim 10^{-13}\ {\rm GeV}^{-2}\left(\frac{\rm GeV}{m_\chi}\right)^2,
\ee
while indirect detection is usually sensitive to pair-annihilation cross sections on the order of $10^{-10}\ {\rm GeV}^{-2}$. Strong constraints from annihilation effects on the CMB would however apply in the case of light dark matter masses.

The cross-symmetric process, $\chi B_1\to \chi B_2$, is instead rather promising, as the implied rates (which again, in general do depend on the underlying model) might be large enough to be of interest for direct detection, provided a modified cosmology affects DM freeze-in pair production.

We postpone a more general and comprehensive analysis of implications of a modified cosmology with a fast-expanding universe at DM freeze-in, including the discussion of specific models, to future studies.

\section{Conclusions}
\label{sec:conclusions}
The cosmological history of the universe is observationally and quantitatively tested only up to temperature of around 1 MeV: at larger temperatures, it is customary to {\it assume} a radiation dominated universe, which is thus the canvas on which pictures for dark matter production in the early universe are usually drawn. However, cosmological histories where at a given temperature the expansion rate, and thus the Hubble rate, was much larger are possible, and yield dramatic consequences for the prediction of the amount of dark matter produced in the early universe by thermal or non-thermal processes.

Here, we focused on the case of dark matter production via {\it freeze-in}: the dark matter is ``dumped'' by a decay or scattering process in the early universe, and never reaches thermal equilibrium. As is well known, given a certain cosmological history, and similarly to the case of thermal freeze-out, it is possible to compute the relic dark matter yield for freeze-in from a few particle physics input parameter characterizing the dark matter sector. Also in analogy to what we recently pointed out for the case of freeze-out in Ref.~\cite{DEramo:2017gpl}, in the presence of a modified cosmological history at temperatures above Big Bang nucleosynthesis, such relic dark matter yield can be profoundly affected, and the ensuing phenomenological and observational consequences for a given dark matter particle setup drastically changed.

To outline a simple yet comprehensive picture of the effects of a rapidly expanding pre-BBN universe, here we parameterized the additional energy density responsible for the modified expansion history with only two parameters, $T_r$ and $n$, effective describing the normalization and the power-law temperature/redshift dependence of the extra species (concrete models for the cosmological history might feature a more complicated functional dependence for the energy density and thus the Hubble rate, see e.g. the recent Ref.~\cite{Dutta:2017fcn}, but the resulting effects fall within the range of parameters we study here).

We focused our study on three specific mechanisms of dark matter freeze-in: (i) the production from decay of some other particle species in the early universe, (ii) the production of a single dark matter particle in the final state of a $2\to2$ scattering process, and (iii) the production of a dark matter pair from a scattering process. For each case, we provided complete expression for the relevant collision operators, reducing the task of calculating the resulting freeze-in abundance to a simple integral. The general and universal finding is that {\it in a faster-than-standard expanding universe, freeze-in production is suppressed}, implying that to produce enough dark matter to match observations, larger couplings, and thus larger detection rates, are in order.

For each of the three cases, we illustrated the freeze-in production suppression, for various values of the parameters defining the cosmological background; We derived analytical expressions that accurately capture and illustrate our numerical results; We then specialized our analysis to simplified expressions for the decay or scattering rates, translating the freeze-in production suppression in the enhancement needed in the relevant particle coupling; Finally, for each of the three cases, we scanned the parameter space of background modified cosmologies, and calculated for each parameter space point the resulting freeze-in production suppression.

Our results are remarkable first for their generality: we demonstrated that in a fast-expanding universe, freeze-in dark matter production is systematically, and dramatically suppressed. Secondly, our results quantify such suppression, which, we find, can be as large as ten orders of magnitude in some cases. Thirdly, and perhaps most importantly, our work outlines the range of potential implications for collider studies and for direct and indirect dark matter detection, which can drastically affect detection strategies for entire classes of particle dark matter candidates.

\section*{Acknowledgments}
This work was supported by the U.S. Department of Energy grant number DE-SC0010107. 

\appendix

\section{Collision Operators}
\label{app:thermalaverages}

The explicit expression for the collision operator $\mathcal{C}_\alpha$ appearing in the Boltzmann equation~\eqref{eq:BEforFI} depends on the specific freeze-in process $\alpha$ under consideration. In this Appendix, we derive its expression for the reactions considered in this work. 

Bath particles always have an equilibrium phase space distribution $f^{\rm eq}_{B_i}(E, t)$ that depends on time and energy, under the assumptions that the universe is homogeneous and isotropic. Equilibrium number densities are defined as follows~\cite{Kolb:1990vq}
\be
n^{\rm eq}_{B_i}(t) =  g_{B_i} \int \frac{d^3 p}{(2\pi)^3} f^{\rm eq}_{B_i}(E_{B_i}(\absval{\bs p}), t) \ .
\label{eq:neqdef} 
\ee
Here, $g_{B_i}$ accounts for internal degrees of freedom (e.g. spin or color) and the dispersion relation reads
\be
E_{B_i}(\absval{\bs p}) = \sqrt{\absval{\bs p}^2 + m_{B_i}^2} \ .
\label{eq:disprel}
\ee 
From now on, we leave the time dependence implicit. In the early universe we are always away from Bose condensation or Fermi degeneracy. This allows us to use $f^{\rm eq}_{B_i}(E_{B_i}) = \exp[ - E_{B_i} / T]$ for both bosons and fermions in thermal equilibrium, and the number density of bath particles reads
\be
n^{\rm eq}_{B_i} = \frac{g_{B_i}}{2 \pi^2} \, m_{B_i}^2 T \, K_2[m_{B_i}/T] \ ,
\label{eq:appneq}
\ee
where $K_2$ is the modified Bessel function. Another useful quantity for the analysis of this Appendix is the Lorentz invariant phase space
\be
d \Pi_{B_i} = \frac{d^3 p_i}{2 E_{B_i} (2\pi)^3} \ .
\label{eq:dPiBi}
\ee

\subsection{Collision Operator for Decays}

We start with the derivation of the collision operator for the decay processes considered in Sec.~\ref{sec:decays}. The number density of $\chi$ can change both due to decays and inverse decays. Here, we only consider decays since DM particles never thermalize and the reaction goes only toward one direction. The collision operator is thus
\be
\begin{split}
\mathcal{C}_{B_1 \rightarrow B_2 \chi} = & \, \int d \Pi_{B_1} \, d \Pi_{B_2} \, d \Pi_\chi \; f^{\rm eq}_{B_1} \,  
\absval{\mathcal{M}_{B_1 \rightarrow B_2 \chi}}^2  \\ & (2 \pi)^4  \delta^4(p_{B_1} - p_{B_2} - p_X) \ .
\end{split}
\ee
The decaying bath particles $B_1$ are in thermal equilibrium. It is important to emphasize here that the squared matrix element in the above equation in summed over initial {\it and} final states. In particular, we do not average over initial polarizations. We identify the {\it partial} decay width for the channel $B_1 \rightarrow B_2 \chi$ computed in the rest frame of $B_1$ and we rewrite the collision operator~\footnote{The partial width $\Gamma_{B_1 \rightarrow B_2 \chi}$ can be different from the total width $\Gamma_{B_1}$ if other decay channels for $B_1$ are allowed.}
\be
\mathcal{C}_{B_1 \rightarrow B_2 \chi} =  g_{B_1}  \, \Gamma_{B_1 \rightarrow B_2 \chi} \,  \int \frac{d^3 p}{(2\pi)^3} \, \frac{m_{B_1}}{E_{B_1}} \, f^{\rm eq}_{B_1}   \ .
\ee
We perform the last integration and we find
\be
\mathcal{C}_{B_1 \rightarrow B_2 \chi} = n_{B_1}^{\rm eq} \, 
\Gamma_{B_1 \rightarrow B_2 \chi}  \, \frac{K_1[m_{B_1}/T]}{K_2[m_{B_1}/T]}  \ ,
\label{eq:appdecay}
\ee
where the equilibrium number density of the decaying bath particle is given in \Eq{eq:appneq}.

\subsection{Collision Operator for Scattering}

The other freeze-in process we consider in this work is production via scattering. As done in Sec.~\ref{sec:scattering}, we distinguish between single and double production.

\subsubsection{Single Production}

For single production the collision operator reads
\be
\begin{split}
& \,  \mathcal{C}^{(a)}_{B_1 B_2 \rightarrow B_3 \chi} = \int d \Pi_{B_1} \, d \Pi_{B_2} \, d \Pi_{B_3} \, d \Pi_\chi \; 
f^{\rm eq}_{B_1} \, f^{\rm eq}_{B_2} \, \times   \\ &
\; \absval{\mathcal{M}_{B_1 B_2 \rightarrow B_3 \chi}}^2   (2 \pi)^4  \delta^4(p_{B_1} + p_{B_2} - p_{B_3} - p_\chi) \ .
\end{split}
\label{eq:appcoll1}
\ee
The initial state bath particles are in equilibrium and the squared matrix element is summer over both initial {\it and} final polarizations, without taking any average as before. Before we further develop the expression above, we observe that it can be rewritten into an equivalent form. Conservation of energy enforces the equality 
\be
\begin{split}
f^{\rm eq}_{B_1} \, f^{\rm eq}_{B_2} = & \, \exp[ - (E_{B_1} + E_{B_2})  / T] =  \\ &
\exp[ - (E_{B_3} + E_{\chi})  / T] = f^{\rm eq}_{B_3} \, f^{\rm eq}_{\chi} \ .
\end{split}
\label{eq:appconsenergy}
\ee
Moreover, if we assume CP invariance, we have the equality between the squared matrix elements
\be
\absval{\mathcal{M}_{B_1 B_2 \rightarrow B_3 \chi}}^2 = \absval{\mathcal{M}_{B_3 \chi \rightarrow B_1 B_2}}^2
\ee
Putting these two results together, we have 
\be
\begin{split}
& \,  \mathcal{C}^{(b)}_{B_1 B_2 \rightarrow B_3 \chi} = \int  d \Pi_{B_3} \, d \Pi_\chi \, d \Pi_{B_1} \, d \Pi_{B_2} \; 
f^{\rm eq}_{B_3} \, f^{\rm eq}_{\chi} \, \times   \\ &
\; \absval{\mathcal{M}_{B_3 \chi \rightarrow B_1 B_2}}^2   (2 \pi)^4  \delta^4(p_{B_3} + p_\chi - p_{B_1} - p_{B_2}) \ .
\end{split}
\label{eq:appcoll2}
\ee

The expressions in Eqs.~(\ref{eq:appcoll1}) and (\ref{eq:appcoll2}) are equivalent forms for the collision operator and they give the same result. In spite of $f^{\rm eq}_{\chi}$ appearing in the second one, DM particles never reach thermal equilibrium. Conservation of energy as expressed in \Eq{eq:appconsenergy} brings $f^{\rm eq}_{\chi}$ into the game, but we are still averaging over initial state bath particles. Although the two expressions are equivalent, it is computationally advantageous to use the one for the reaction allowed at zero kinetic energy: in other words, if $m_{B_1} + m_{B_2} > m_{B_3} + m_{\chi}$ we use \Eq{eq:appcoll1}, otherwise \Eq{eq:appcoll2}. This strategy isolates thermal suppressions in the distribution functions rather than phase space integrals. In what follows, we develop both expressions. 

We present the derivation starting from \Eq{eq:appcoll1}; the one correspondent to the definition in \Eq{eq:appcoll2} is analogous. We define the Lorentz invariant relative velocity between the two initial state particles 
\be
v_{B_1 B_2} \equiv \dfrac{\sqrt{(p_{B_1} \cdot p_{B_2})^2- m_{B_1}^{2}m_{B_2}^{2}}}{p_{B_1} \cdot p_{B_2}} \ .
\ee
Here, $p_{B_i}$ (with $i = 1, 2$) are Lorentz four-vectors denoting initial state four-momenta, and the only consider Lorentz invariant products. Once we put particles on-shell ($p_{B_i}^2 = m_{B_i}^2$), the relative velocity reads
\be
v_{B_1 B_2} = \frac{\lambda^{1/2}(s, m_{B_1}, m_{B_2})}{2\; p_{B_1}\cdot p_{B_2}} \ ,
\label{eq:appvB1B2}
\ee
where we introduce the (square of the) center of mass energy $s = \left( p_{B_1} + p_{B_2} \right)^2$ and we define the function
\be
\lambda(x, y, z) \equiv [x - (y + z)^2] [x - (y - z)^2]  \ .
\label{eq:lambdadef}
\ee

The Lorentz invariant cross section for each individual binary collision is defined as it follows~\cite{Peskin:1995ev}
\be
\begin{split}
& \, \sigma_{B_1 B_2 \rightarrow B_3 \chi}(s) = \frac{1}{g_{B_1} g_{B_2}} \frac{1}{4 \, p_{B_1}\cdot p_{B_2} \, v_{B_1 B_2} } \\ & \int d \Pi_{B_3} \, d \Pi_\chi  \, \absval{\mathcal{M}_{B_1 B_2 \rightarrow B_3 \chi}}^2   \\ & (2 \pi)^4  \delta^4(p_{B_1} + p_{B_2} - p_{B_3} - p_\chi) \ .
\end{split}
\label{eq:appbinaryXS}
\ee
According to our conventions, the squared matrix element appearing in \Eq{eq:appcoll1} is only summed over initial states, and this is why we divided the expression above by an overall factor of $g_{B_1} g_{B_2}$. This allows us to express the collision operator in \Eq{eq:appcoll1} in terms of a thermally averaged cross section
\be
\begin{split}
\mathcal{C}^{(a)}_{B_1 B_2 \rightarrow B_3 \chi} = & \, 2 g_{B_1} g_{B_2} \int d \Pi_{B_1} \, d \Pi_{B_2} \, f^{\rm eq}_{B_1} \, f^{\rm eq}_{B_2} \\ & \lambda^{1/2}(s, m_{B_1}, m_{B_2}) \sigma_{B_1 B_2 \rightarrow B_3 \chi}(s) \ ,
\end{split}
\label{eq:appcoll3}
\ee
where we use \Eq{eq:appvB1B2} for the relative velocity.

The last task left for us is the phase space integration. The integrand depends only on the energies $E_{B_1}$ and $E_{B_2}$ and on $s$, thus the only non-trivial angular integration is the one over the angle $\theta$ between the initial momenta. The integration over the remaining  angles is straightforward. After plugging in the definition in \Eq{eq:dPiBi}, the integration measure reads
\be
\begin{split}
d \Pi_{B_1} \, d \Pi_{B_2} = & \, 
\frac{|p_{B_1}|^2 \, d|p_{B_1}| \, d\Omega_{B_1}}{16 \pi^3 E_{B_1}}  
\frac{|p_{B_2}|^2 \, d|p_{B_2}| \, d\Omega_{B_2}}{16 \pi^3 E_{B_2}} = \\ &
 \frac{|p_{B_1}| \, |p_{B_2}|}{32 \pi^4} \, dE_{B_1} dE_{B_2} d\cos\theta  \ ,
\end{split}
\ee
where in the second row we perform the straightforward integration over the angles and we we use the dispersion relation in \Eq{eq:disprel}. In order to proceed, it is convenient to use the following variables~\cite{Gondolo:1990dk}
\begin{align}
E_{+} = & \, E_{B_1} + E_{B_2} \ , \\
E_{-} = & \, E_{B_1} - E_{B_2} \ , \\ \nonumber 
s = & \, m_{B_1}^{2} + m_{B_2}^{2} + \\ & 
\label{eq:costheta}  2 \left( E_{B_1}E_{B_2} - |p_{B_1}| |p_{B_2}| \cos \theta \right) \ .
\end{align}
The Jacobian for this transformation reads
\be
 d E_{B_1} d E_{B_2} d \cos\theta = \frac{dE_+ dE_- ds}{4 |p_{B_1}| |p_{B_2}|} \ ,
\ee
and the integration measure expressed in terms of the new variables takes a much simpler form
\be
d \Pi_{B_1} \, d \Pi_{B_2} = \frac{dE_+ dE_- ds}{128 \pi^4}  \ .
\ee

Before computing the integral, we need to identify the integration domain. The Mandelstam variables $s$ is bound to be in the region
\be
s \geq s^{\rm min}_{12} \equiv (m_{B_1} + m_{B_2})^2 \ .
\ee
Once we fix $s$, the variable $E_+$ can take the values
\be
E_+ = \sqrt{s - \left(\bs p_{B_1} + \bs p_{B_2} \right)^2}  \geq \sqrt{s} \ .
\ee
The allowed values for $E_{-}$ are found after imposing that the absolute value of $\cos\theta$ as expressed in \Eq{eq:costheta} is always smaller than one. We find the range
\be
\frac{\left| E_- - E_+ \frac{(m^{2}_{B_1}-m^{2}_{B_2})}{s} \right|}{\left(E_+^2 - s\right)^{1/2}} \leq \frac{\lambda^{1/2}(s,m_{B_1},m_{B_2})}{s} \ .
\ee

Finally, we perform the integrations. The product $f^{\rm eq}_{B_1} \, f^{\rm eq}_{B_2} =  \exp[ - E_+ / T]$ depends only on $E_+$, therefore we can always perform the integration over $d E_-$
\be
\begin{split}
& \mathcal{C}^{(a)}_{B_1 B_2 \rightarrow B_3 \chi} = \frac{g_{B_1} g_{B_2}}{32 \pi^4}  \, \times \\ &
\int_{s^{\rm min}_{12}}^\infty ds \, \frac{\lambda(s, m_{B_1}, m_{B_2})}{s} \, \sigma_{B_1 B_2 \rightarrow B_3 \chi}(s)  \\ &
\int_{\sqrt{s}}^\infty dE_+ \exp[ - E_+ / T]  \left(E_+^2 - s\right)^{1/2}   \ .
\end{split}
\label{eq:appcoll4}
\ee
The integral over $dE_+$ gives a Bessel function
\be
\begin{split}
& \mathcal{C}^{(a)}_{B_1 B_2 \rightarrow B_3 \chi} = \frac{g_{B_1} g_{B_2}}{32 \pi^4}  \, T \, \times \\ &
\int_{s^{\rm min}_{12}}^\infty ds \, \frac{\lambda(s, m_{B_1}, m_{B_2})}{s^{1/2}} \, \sigma_{B_1 B_2 \rightarrow B_3 \chi}(s) \, K_1[\sqrt{s} / T] \ .
\end{split}
\label{eq:appcoll5}
\ee
This is our final expression. The last integral over $s$ can be performed only after we specify the explicit cross section, and it is in general model dependent. 

We conclude with two additional results. First, we quote the final expression for the collision operator as defined in \Eq{eq:appcoll2}. After a similar derivation to the decay case, we find
\be
\begin{split}
& \mathcal{C}^{(b)}_{B_1 B_2 \rightarrow B_3 \chi} = \frac{g_{B_3} g_{\chi}}{32 \pi^4}  \, T \, \times \\ &
\int_{s^{\rm min}_{3\chi}}^\infty ds \, \frac{\lambda(s, m_{B_3}, m_{\chi})}{s^{1/2}} \, \sigma_{B_3 \chi \rightarrow B_1 B_2}(s) \, K_1[\sqrt{s} / T] \ ,
\end{split}
\label{eq:appcoll6}
\ee
where this time $s^{\rm min}_{3\chi} = (m_{B_3} + m_\chi)^2$. Second, we introduce a compact form to express the collision operator
\begin{align}
\label{eq:appcoll3averagecross} \mathcal{C}^{(a)}_{B_1 B_2 \rightarrow B_3 \chi} =  & \, \langle \sigma_{B_1 B_2 \rightarrow B_3 \chi} v \rangle \,  n^{\rm eq}_{B_1} n^{\rm eq}_{B_2}  \ , \\
\label{eq:appcoll3averagecross2} \mathcal{C}^{(b)}_{B_1 B_2 \rightarrow B_3 \chi} =  & \, \langle \sigma_{B_3 \chi \rightarrow B_1 B_2} v \rangle \,  n^{\rm eq}_{B_3} n^{\rm eq}_{\chi} \ ,
\end{align}
as a combination of equilibrium number densities and a thermally averaged cross section. The explicit forms for the latter can be obtained by  identifying the equilibrium distribution as defined in \Eq{eq:appneq}, and they result in
\begin{align}
& \nonumber \langle \sigma_{B_1 B_2 \rightarrow B_3 \chi} v \rangle =  \frac{1}{8 \, K_2[m_{B_1} / T] K_2[m_{B_2} / T] \, m_{B_1}^2 m_{B_2}^2 T} \\ &
\label{eq:thermaverag1} \int_{s^{\rm min}_{12}}^\infty ds \frac{\lambda(s, m_{B_1}, m_{B_2})}{s^{1/2}} \sigma_{B_1 B_2 \rightarrow B_3 \chi}(s)  K_1[\sqrt{s} / T]  \ . \\
& \nonumber \langle \sigma_{B_3 \chi \rightarrow B_1 B_2} v \rangle =  \frac{1}{8 \, K_2[m_{B_3} / T] K_2[m_\chi / T] \, m_{B_3}^2 m_{\chi}^2 T} \\ &
\label{eq:thermaverag2}  \int_{s^{\rm min}_{3\chi}}^\infty ds \frac{\lambda(s, m_{B_3}, m_{\chi})}{s^{1/2}} \sigma_{B_3 \chi \rightarrow B_1 B_2}(s)  K_1[\sqrt{s} / T] \ .
\end{align}
The equality between the collision operators expressed as in Eqs.~(\ref{eq:appcoll1}) and (\ref{eq:appcoll2}) can be also written as
\be
\langle \sigma_{B_1 B_2 \rightarrow B_3 \chi} v \rangle \,  n^{\rm eq}_{B_1} n^{\rm eq}_{B_2}   = \langle \sigma_{B_3 \chi \rightarrow B_1 B_2} v \rangle \,  n^{\rm eq}_{B_3} n^{\rm eq}_{\chi} \ .
\ee

\subsubsection{Pair Production}

The collision operator for the case of DM pair production can be derived by employing similar techniques. As usual, the collision operator can be written in two equivalent forms. Here, we report the final results
\begin{align}
& \nonumber  \mathcal{C}^{(a)}_{B_1 B_2 \rightarrow \chi \chi} = \frac{g_{B_1} g_{B_2}}{32 \pi^4}  \, T \, \times \\ &
\label{eq:appcollpair1} \int_{s^{\rm min}_{12}}^\infty ds \, \frac{\lambda(s, m_{B_1}, m_{B_2})}{s^{1/2}} \, \sigma_{B_1 B_2 \rightarrow \chi \chi}(s) \, K_1[\sqrt{s} / T] \ . \\
& \nonumber \mathcal{C}^{(b)}_{B_1 B_2 \rightarrow \chi \chi} = \frac{g^2_{\chi}}{32 \pi^4}  \, T \, \times \\ &
\label{eq:appcollpair2} \int_{s^{\rm min}_{\chi\chi}}^\infty ds \, \frac{\lambda(s, m_{\chi}, m_{\chi})}{s^{1/2}} \, \sigma_{\chi \chi \rightarrow B_1 B_2}(s) \, K_1[\sqrt{s} / T] \ .
\end{align}
As already done before, we give expressions for both cases of direct and inverse reactions. We can also write the collision operators in the form
\begin{align}
\mathcal{C}^{(a)}_{B_1 B_2 \rightarrow \chi \chi} =  & \, \langle \sigma_{B_1 B_2 \rightarrow \chi \chi} v \rangle \,  n^{\rm eq}_{B_1} n^{\rm eq}_{B_2} \ , \\
\mathcal{C}^{(b)}_{B_1 B_2 \rightarrow \chi \chi} =  & \, \langle \sigma_{\chi \chi \rightarrow B_1 B_2} v \rangle \,  n^{\rm eq}_{\chi} n^{\rm eq}_{\chi}  \ ,
\end{align}
where the thermally averaged cross sections result in
\begin{align}
& \nonumber \langle \sigma_{B_1 B_2 \rightarrow \chi \chi} v \rangle =  \frac{1}{8 \, K_2[m_{B_1} / T] K_2[m_{B_2} / T] \, m_{B_1}^2 m_{B_2}^2 T} \\ & \label{eq:appcollscattxxcross}
\int_{s^{\rm min}_{12}}^\infty ds \frac{\lambda(s, m_{B_1}, m_{B_2})}{s^{1/2}} \sigma_{B_1 B_2 \rightarrow \chi \chi}(s)  K_1[\sqrt{s} / T ] \ ,\\
& \nonumber \langle \sigma_{\chi \chi \rightarrow B_1 B_2} v \rangle =  \frac{1}{8 \, K_2[m_{\chi} / T]^2 \, m_{\chi}^4 T} \\ &
\int_{s^{\rm min}_{\chi \chi}}^\infty ds \frac{\lambda(s, m_{\chi}, m_{\chi})}{s^{1/2}} \sigma_{\chi \chi \rightarrow B_1 B_2}(s)  K_1[\sqrt{s} / T ] \ .
\end{align}

\subsubsection{Some Limiting Expressions}

All results derived in this Appendix so far did not rely upon any approximation. Here, we conclude by providing some limiting expressions that are useful for the analytical estimates found in this work. The scattering analysis in Sec.~\ref{sec:scattering} always assumes a constant matrix element for the collision. In other words, we always consider matrix element independent on the kinematics. Within this assumption, the cross section for binary collisions in \Eq{eq:appbinaryXS} can be immediately computed because the phase space integral is straightforward. 

For single DM production, and within this assumption, the binary cross section reads
\be
\sigma_{B_1 B_2 \rightarrow B_3 \chi}(s) = \frac{\absval{\mathcal{M}_{B_1 B_2 \rightarrow B_3 \chi}}^2}{g_{B_1} g_{B_2} \, 16 \pi s} \frac{\lambda^{1/2}(s, m_{B_3}, m_{\chi})}{\lambda^{1/2}(s, m_{B_1}, m_{B_2})} \ .
\ee
Likewise, the cross section for the inverse reaction reads
\be
\sigma_{B_3 \chi \rightarrow B_1 B_2}(s) = \frac{\absval{\mathcal{M}_{B_3 \chi \rightarrow B_1 B_2}}^2}{g_{B_3} g_{\chi}\, 16 \pi s} \frac{\lambda^{1/2}(s, m_{B_1}, m_{B_2})}{\lambda^{1/2}(s, m_{B_3}, m_{\chi})} \ .
\ee
The collision operator can be computed from Eqs.~\eqref{eq:appcoll5} or \eqref{eq:appcoll6}. Both expressions give the same result
\be
\begin{split}
& \mathcal{C}^{(a)}_{B_1 B_2 \rightarrow B_3 \chi} =  \mathcal{C}^{(b)}_{B_1 B_2 \rightarrow B_3 \chi} = \\ &
\frac{\absval{\mathcal{M}_{B_1 B_2 \rightarrow B_3 \chi}}^2 \, T}{512 \pi^5} \,  \int_{s_{\rm single}^{\rm min}}^\infty \frac{ds}{s^{3/2}} \, K_1[\sqrt{s} / T] \, \times \\ &  \lambda^{1/2}(s, m_{B_1}, m_{B_2}) \lambda^{1/2}(s, m_{B_3}, m_{\chi}) \ ,
\end{split}
\label{eq:appcollfinal1}
\ee
where the lower integration limit is set by the kinematical threshold for the reaction $s_{\rm single}^{\rm min} = {\rm max}\left\{s^{\rm min}_{12} , s^{\rm min}_{3\chi} \right\}$. The remaining integral depends on the spectrum of the model and it can be computed numerically. 

For DM pair production, an analogous calculation leads to the result
\be
\begin{split}
& \mathcal{C}^{(a)}_{B_1 B_2 \rightarrow \chi \chi} =  \mathcal{C}^{(b)}_{B_1 B_2 \rightarrow \chi \chi} = \\ &
\frac{\absval{\mathcal{M}_{B_1 B_2 \rightarrow \chi \chi}}^2 \, T}{512 \pi^5} \,  \int_{s_{\rm pair}^{\rm min}}^\infty \frac{ds}{s^{3/2}} \, K_1[\sqrt{s} / T] \, \times \\ &  \lambda^{1/2}(s, m_{B_1}, m_{B_2}) \lambda^{1/2}(s, m_{\chi}, m_{\chi}) \ ,
\end{split}
\label{eq:appcollfinal2}
\ee
where this time $s_{\rm pair}^{\rm min} = {\rm max}\left\{s^{\rm min}_{12} , s^{\rm min}_{\chi\chi} \right\} $.

\bibliography{EFCpaper2}

\end{document}